\documentclass[pra,twocolumn,letterpaper,showpacs,groupedaddress,superscriptaddress,nofootinbib,floatfix,preprintnumbers,tightenlines]{revtex4-1}
\usepackage{hyperref,amssymb,amsmath,amsfonts,graphicx,xcolor, mathtools,amscd,amsthm}
\usepackage{bm}
\usepackage{verbatim}
\usepackage{url}
\usepackage[braket, qm]{qcircuit}
\usepackage[section]{placeins}
\usepackage{xcolor}

\def\Dslash{D\hskip-0.65em /}

\def\OMIT#1{{}}
\def\Dslash{D\hskip-0.65em /}

\def\Tslash{{T\hskip-0.48em /}}

\newcommand{\expect}[1]{\langle {#1} \rangle}

\begin{document}

%\title{Quantum-Classical Dynamics of the Schwinger Model using Quantum Computers}
\title{Quantum-Classical Computation of Schwinger Model Dynamics using Quantum Computers}

\author{N. Klco}
\email[email: ]{klcon@uw.edu}
\affiliation{Institute for Nuclear Theory, University of Washington, Seattle, WA 98195-1550, USA}

\author{E. F. Dumitrescu}
\affiliation{Computational Sciences and Engineering Division, Oak Ridge National Laboratory,
  Oak Ridge, TN 37831, USA}

\author{A. J. McCaskey}
\affiliation{Computer Science and Mathematics Division, Oak Ridge National Laboratory,
  Oak Ridge, TN 37831, USA}

\author{T.~D.~Morris}
\affiliation{Physics Division, Oak Ridge National Laboratory,
Oak Ridge, TN 37831, USA}

\author{R.~C.~Pooser}
\affiliation{Computational Sciences and Engineering Division, Oak Ridge National Laboratory,
Oak Ridge, TN 37831, USA}

\author{M. Sanz}
\affiliation{Department of Physical Chemistry, University of the Basque Country UPV/EHU, Apartado 644, E-48080 Bilbao, Spain.}

\author{E. Solano}
\affiliation{Department of Physical Chemistry, University of the Basque Country UPV/EHU, Apartado 644, E-48080 Bilbao, Spain.}
\affiliation{IKERBASQUE, Basque Foundation for Science, Maria Diaz de Haro 3, E-48013 Bilbao, Spain}

\author{P. Lougovski}
\email[email: ]{lougovskip@ornl.gov}
\affiliation{Computational Sciences and Engineering Division, Oak Ridge National Laboratory,
Oak Ridge, TN 37831, USA}

\author{M. J. Savage}
\email[email: ]{mjs5@uw.edu}
\affiliation{Institute for Nuclear Theory, University of Washington, Seattle, WA 98195-1550, USA}

\date{\today}

\preprint{INT-PUB-18-013}

%%%%%%%%%%%%%%%%%%%%%%%%%%%%%%%%%%%%%%%%%%%%%%
\begin{abstract}
We present a quantum-classical algorithm to study the dynamics of the two-spatial-site Schwinger model on IBM's quantum computers. Using rotational symmetries, total charge, and parity, the number of qubits needed to perform computation is reduced by a factor of $\sim 5$, removing exponentially-large unphysical sectors from the Hilbert space. Our work opens an avenue for exploration of other lattice quantum field theories, such as quantum chromodynamics,  where classical computation is used to find symmetry sectors in which the quantum computer evaluates the dynamics of quantum fluctuations.
\end{abstract}

\maketitle

%%%%%%%%%%%%%%%%%%%%%%%%%%%%%%%%%%%%%%%%%%%%%%%

\section{Introduction}
Quantum field theories (QFTs),
and in particular gauge field theories,
provide the mathematical framework to describe three of the four fundamental forces of nature.
 In quantum chromodynamics (QCD),
 the gauge theory describing the strong interactions~\cite{Fritzsch:1973pi,Politzer:1973fx,Gross:1973id}, the invariance of  the laws of nature under ${\rm SU(3)_c}$ transformations necessitate the existence of
 eight gluon fields that transmit the forces between the quarks.
When calculating QCD phenomena in the high energy (short distance) limit, perturbative techniques, such as Feynman diagram expansions, is efficacious.
However, difficulties arise in applying such approaches to  low-energy processes, in which color confinement and the spontaneous breaking of approximate chiral symmetries  dominate  structure and dynamics.
 This regime requires
the use of low-energy effective field theories, such as chiral perturbation theory ($\chi$PT)~\cite{Weinberg:1968de}, and  numerical solutions using Lattice QCD (LQCD)~\cite{Wilson:1974sk}.
Exascale classical computing will address Grand Challenge problems~\cite{ASCRexascale} in nuclear and high-energy physics by enabling high-precision LQCD calculations of many properties of hadrons and light nuclei as well as  low-energy scattering processes. However, these resources are likely insufficient to address other questions and problems of importance, such as the structure, properties and dynamics of finite-density systems (due to the presence of sign problems in the algorithms used on conventional computers) or the fragmentation of high energy quarks and gluons into hadrons.
 Quantum computers may offer potential solutions in these systems that are inaccessible with conventional computing~\cite{Lloyd1996,OrtizPhysRevA2001,SommaPhysRevA2002,Byrnes:2005qx,JordanLeePreskill2012,Jordan:2011ci,Zohar:2012xf,Zohar:2012ay,Banerjee:2012pg,Banerjee:2012xg,Wiese:2013uua,Jordan:2014tma,Marcos:2014lda,Wiese:2014rla,Zohar:2016iic,Jordan:2017lea,Bermudez:2017yrq}.

 Existing and near-term quantum hardware is imperfect, with a small number of qubits, sparse qubit connectivity, and noisy quantum gates---all hallmarks of quantum computers in the NISQ (Noisy Intermediate-Scale Quantum) era~\cite{1801.00862}. These technical imperfections constrain the circuit depth and dimensionality of problems that can be solved on available quantum computers. Nonetheless, recent advances in developing~\cite{JordanLeePreskill2012,Wiese:2013uua,Wiese:2014rla,GarciaAlvarezFermionFermionPRL2015,zohar2016,Pichler2016,Bermudez:2017yrq,PhysRevLett.121.110504} and implementing~\cite{Zohar:2012xf,Zohar:2012ay,Banerjee:2012xg,Banerjee:2012pg,Marcos:2014lda,Zohar:2016iic,Martinez2016} quantum algorithms for QFT calculations have improved our understanding of the algorithmic complexity of the problem. On the other hand, rapid progress in quantum simulations of many-body systems, such as molecules and spin chains~\cite{lanyon2010,Peruzzo2014,omalley2016,kandala2017}, has mapped out potential ways to reduce complexity through combinations of classical and quantum computation methods, with variational approaches~\cite{McClean2016,Santagati2018} at the forefront of new developments.

In this work, we develop a hybrid quantum-classical computation strategy for a prototypical lattice gauge QFT---the Schwinger 1+1 dimensional model~\cite{Schwinger:1962tp,Coleman:1975pw} on the lattice. Using this strategy, we study the ground state properties as well as the real-time dynamics of particle and electric field energy density.
In contrast to previous works~\cite{Martinez2016,Muschik2017},
we employ periodic boundary conditions (PBCs) endowing the lattice with discrete rotational symmetries and reflection symmetries.
Projections into symmetry sectors
lead to a refined classification of states in the Hilbert space by their momentum, charge and parity (projections used in LQCD calculations).  This leads to a significant reduction of the Hilbert space of the system, confining calculations to physically allowed states. The task of determining the physical sectors of the Hilbert space is outsourced to classical computers. The dynamics of the model within each symmetry sector are evaluated using a digital quantum computer by applying unitary operators and implementing them as a sequence of one-qubit and two-qubit gates. As an exploration of what is currently practical on state-of-the-art quantum computers, we solve for the dynamics of the Schwinger model with one and two spatial lattice sites using IBM's quantum computer.

\section{The Schwinger Model}
\begin{figure}[!ht]
	\includegraphics[width=0.8\columnwidth]{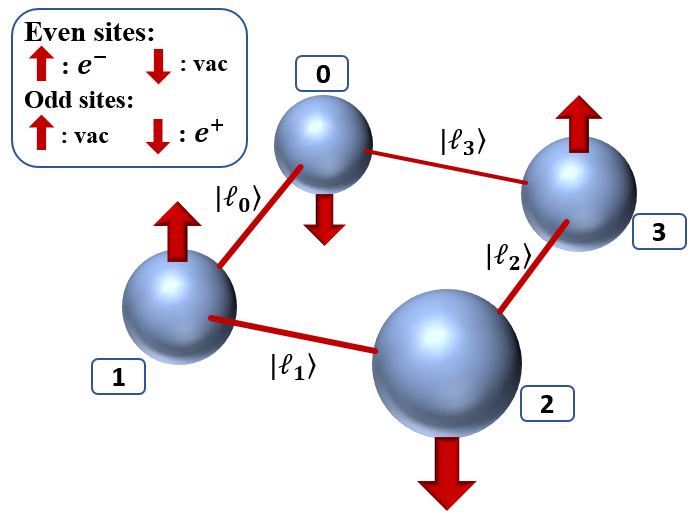}
	\caption{
	A schematic of the qubit and electric flux link structure of the two-spatial-site lattice Schwinger model. Even sites (marked 0 and 2) represent the electron content with spin up denoting the presence of an electron.  Odd sites (marked 1 and 3) represent the positron content with spin down denoting the presence of a positron. The strong-coupling vacuum (unoccupied) state is antiferromagnetic.
	}
	\label{fig:zp1}
\end{figure}

The Schwinger model describes quantum electrodynamics in one space and one time dimension.
It enjoyed popularity in the 1960's and 1970's as a ``prototype'' for the strong interactions as it shares with QCD a number of features, such as confinement and spontaneous breaking of chiral symmetry.
After gauge-fixing, there is only one dynamical component of the photon field,
which acquires a mass through quantum fluctuations.
Charge is screened, the lightest excitation in the spectrum has the quantum numbers of the photon, and
the vacuum of the theory enjoys a non-zero condensate, $\langle\psi\overline{\psi}\rangle$.
The Lagrange density that defines the continuum Schwinger model,
\begin{eqnarray}
{\cal L} & = &
\overline{\psi}\left(i\Dslash - m\right)\psi - {1\over 4}F_{\mu\nu}F^{\mu\nu}
\ \ \ ,
\end{eqnarray}
can be spatially discretized with the Kogut-Susskind (staggered) action~\cite{KogutSusskind1975,Banks:1975gq,Casher:1973uf},
 mapped onto a (re-scaled)
Hamiltonian density using the Jordan-Wigner transformation, and gauge-fixed by setting the temporal component of the gauge field to zero ($A_0=0$) on $N_{fs}/2$ spatial sites,
\begin{eqnarray}
\hat H & = &
x \sum_{n=0}^{N_{fs}-1} \left( \sigma_n^+ L_n^- \sigma^-_{n+1}  + \sigma_{n+1}^+ L_n^+ \sigma^-_{n}\right) \nonumber \\
& + & \sum_{n=0}^{N_{fs}-1} \left(l_n^2  +  {\mu\over 2} (-)^n \sigma_n^z\right)
\ \ .
\label{eq:KSham}
\end{eqnarray}
This Kogut-Susskind action distributes the two components of the 1-dimensional fermion field across neighboring even-odd sites and results in two fermion sites per spatial site (see Figure~\ref{fig:zp1} for a two-spatial-site example).
The first term in $\hat H$ corresponds to the kinetic energy of the fermion field (a hopping term), the second term is the total energy in the electric field, and the third term is the mass term.
The couplings in Eq.~(\ref{eq:KSham}) are related to the value of the gauge coupling $g$, the lattice spacing $a$
and the fermion mass $m$,
$x=1/(ag)^2$ and $\mu= 2m/(a g^2)$.
The $l_n$'s are integers, ranging between $-\infty$ and $+\infty$, describing the quantized electric flux
in the link between the site $n$ and $n+1$, while the $L_n^\pm$ are link lowering and raising operators acting as $L^\pm |l\rangle = |l\pm1\rangle$.
Two qubits are sufficient to describe the fermion occupation of a single spatial lattice site, one for the $e^-$ and one the $e^+$.
As low energy observables become insensitive to high-energy modes,  the  impact of the necessary ultraviolet
cutoff on each $l_n$ can be quantified and removed~\cite{tHooft:1972tcz,Wilson:1973jj,Appelquist:1974tg,Weinberg:1978kz,Symanzik:1983dc,Kuhn:2014rha}.
While following naturally in Lagrangian dynamics as a Lagrange multiplier, the Gauss's Law constraint relating the electric flux entering and leaving a closed surface to the electric charge contained in that surface must be imposed ``by hand'' in the initial state of a Hamiltonian formulation.
Approaching the strong coupling limit, in which $x,\mu \rightarrow 0$ with their ratio finite or simply $x\rightarrow 0$ for the massless case~\cite{Banks:1975gq}, the vacuum of the theory is perturbatively close to an anti-ferromagnetic phase
with the $e^-$ and $e^+$ qubits anti-aligned (see Fig.~\ref{fig:zp1},
without energy in the electric field).

Recent studies of the dynamic properties, such as charge fluctuations, entanglement entropy evolution, string breaking, and meson scattering in the Schwinger model
have been performed
in trapped ion systems~\cite{Muschik2017,Martinez2016}
or with classical tensor networks~\cite{Pichler2016,Banuls2016,PhysRevD.94.085018,PhysRevD.96.114501}.
In the former, open boundary conditions with vanishing background field are used to truncate the gauge-field Hilbert space.  Constraining the remaining non-dynamical links to satisfy Gauss's law results in long-range, two-body interactions that are feasible with trapped-ion-specific M{\o}lmer-S{\o}rensen gates, but are more severely burdensome in superconducting-circuit quantum computers.  Our work enriches the current literature by retaining local interactions while removing from the calculation not only the exponentially-large, unphysical subspace but also the symmetry-sector-distinct regions of Hilbert space.  As a result, inevitable errors occurring in today's noisy quantum systems are incapable of populating states outside of the correct, dynamical Hilbert space.

Considering first the theory with one spatial site, denoted as $0+1$, the dynamical degrees of freedom are two fermion sites ($N_{fs} = 2$),
the electron and positron occupations, and two flux links.  This system can be visualized as half of the $1+1$ system system with two spatial sites  shown schematically in Fig.~\ref{fig:zp1}.
Though there are many options in regulating the formally-infinite energy of the electric field, we choose to impose a cutoff on the energy in each electric flux link,
$|l_n|\le 1$ (this structure is reminiscent of U(1) quantum link models as discussed in \cite{Wiese:2013uua}).  Increasing this cutoff increases the physical Hilbert space dimension linearly and thus logarithmically increases qubit requirements.
With two quantum states per fermion site and three per flux link, this system contains a total of 36 quantum states that naturally embed in the larger space of six qubits, one for each fermion and two for each link.  These 36 states describe all sectors with charge $Q=0,\pm 1$, with only a subset satisfying Gauss's law.
Working in the $Q=0$ sector, which can be connected to the strong-coupling ground state,
reduces the number of states from 36 down to  5 (see Appendix~\ref{app:MomParCC}).
Taking note of the utility of discrete space-time symmetries in nuclear and particle physics, we consider the transformation of these 5 states under the operation of $\hat P$, parity as defined by the symmetries of the staggered circle.
The parity transformations reflect the system about axes that pass through either two electron or two positron sites.
The 5 physical states are further classified into 3 $P=+1$ states
 and 2 $P=-1$ states.
The quantum evolution of the $P=+1$ sector can be calculated
using two qubits while that of the $P=-1$ sector using one qubit, thereby reducing the required
number of qubits from 6 to 2.
In these sectors, the Hamiltonians take the form,
\begin{equation}
{\rm H}_+ \!=\!
\left(
\begin{array}{lll}
-\mu  & \sqrt{2} x & 0\\
\sqrt{2} x  & 1+\mu & x \\
0  & x & 2-\mu
 \end{array}
 \!\!\right),
{\rm H}_- \!=\!
 \left(
\begin{array}{ll}
1+\mu & x \\
x & 2-\mu
 \end{array}
 \!\!\right)\!\!.
\end{equation}

With two spatial sites, the state reduction procedure parallels that of the one-site theory.
With an energy cutoff of $|l_n|\le 1$ for each link, the 4 fermion sites and 4 flux links
support
a total number of 1296 quantum states contained in 12 qubits---a lattice-inspired implementation on a quantum computer with nearly 99.7\% of the Hilbert space unphysical.
Imposing the Gauss's Law constraint isolates the 13 physical states with $Q=0$ (see Appendix~\ref{app:MomParCC}).
These states can be projected against momentum.
This corresponds to rotating the system by two (of the four) fermion sites and
multiplying by a complex phase, $e^{-i{\bf k}\cdot {\bf x}}$ where ${\bf k}$ corresponds to an
allowed momentum.
The 13 states decompose into sectors defined by momentum,
${\bf k}={\bf 0}, \pm{\bf 1}$
with 9 states residing in the ${\bf k}={\bf 0}$ sector, which contains the vacuum.
The states in the ${\bf k}={\bf 0}$  sector can be further classified with respect to $\hat P$,
providing a 5-state $P=+1$ sector and a 4-state $P=-1$ sector.
For non-zero momentum, $\hat P$ transforms between states of opposite momentum,
creating energy degeneracies between the momentum sectors.  The Hamiltonians in these sectors are
\begin{eqnarray}
{\rm H}_{{\bf k}={\bf 0},+}  & = &
\left(
\begin{array}{ccccc}
-2\mu & 2x & 0 & 0 & 0 \\
2x & 1 &  \sqrt{2} x  & 0 & 0 \\
0 &  \sqrt{2} x  & 2+2\mu &  \sqrt{2} x & 0 \\
0 & 0 &  \sqrt{2} x  & 3 &  \sqrt{2} x  \\
0 & 0 & 0 &  \sqrt{2} x  & 4-2\mu
 \end{array}
 \right),
 \nonumber\\
 {\rm H}_{{\bf k}={\bf 0},-}  & = &
\left(
\begin{array}{cccc}
1 & \sqrt{2} x  & 0 & 0 \\
\sqrt{2} x &  2 + 2\mu & - \sqrt{2} x & 0 \\
 0 & -\sqrt{2} x  & 3  & \sqrt{2} x \\
 0 & 0 &   \sqrt{2} x & 4- 2\mu
 \end{array}
 \right)
\ \ \ ,
\label{eq:H1p1mats}
 \end{eqnarray}
and
$H_{{\bf k}=\pm{\bf 1}} = {\rm diag}(1,3)$,
for which the nearest-neighbor interactions give rise to the band diagonal structure.
The na\"{i}ve requirement of 12 qubits to describe this field theory has been reduced  to 3.
The matrices in Eq.~(\ref{eq:H1p1mats}) are organized in ascending total energy in the electric field.
As the low-energy properties and dynamics of this system will become increasingly insensitive to contributions from
high energy states, a further truncation can be made in which the total energy in the electric field is less than a second cutoff, $\tilde{\Lambda}$.
To contain the ${\bf k}={\bf 0}$ $P=+1$ sector in two qubits, a cutoff of $\sum\limits_n l_n^2 \le 3 = \tilde{\Lambda}$ is imposed, which introduces a systematic error at the $\sim 1\%$-level in the low-lying energies for $x = 0.6$ and $\mu = 0.1$ (see Appendix~\ref{app:exact2site} and~\ref{app:convergence}).
It is important to note that these state reductions were accomplished with classical computing resources. The states comprising symmetry subspaces and Hamiltonian matrix elements over those subspaces were calculated using a classical computer. As can be seen in Tab.~\ref{tab:scaling} of Appendix~\ref{app:largerlattices}, these symmetry-projected Hamiltonian matrix elements require evaluations in an exponentially-growing Hilbert space.
To explore systems larger than those that can be stored on a classical computer, it will be necessary to develop quantum algorithms to accomplish such reductions {\it in situ}.

\section{Ground State Calculations}
\begin{figure}[!t]
    \centering
    \includegraphics[width=\columnwidth]{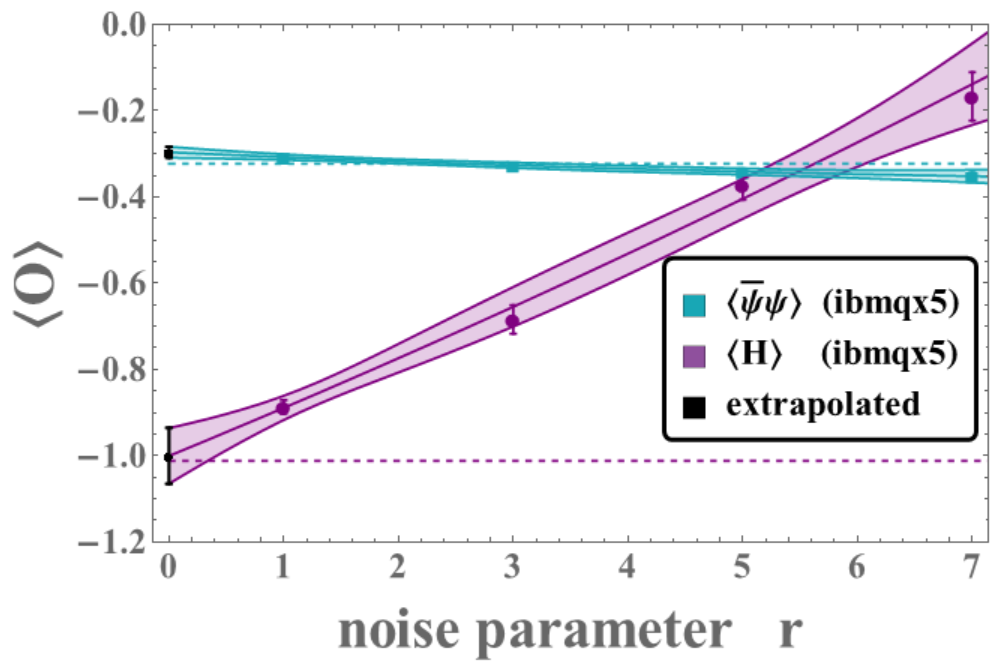}
    \caption{The $H_{\textbf{k=0},+}^{\tilde{\Lambda}=3}$ ground state energy and chiral condensate (purple, blue extrapolated to -1.000(65) and -0.296(13), respectively) expectation values as a function of $r$, the noise parameter.
    $r-1$ is the number of additional CNOT gates inserted at each location of a CNOT gate in the original VQE circuit. (1200 IBM allocation units and $\sim6.4$~QPU$\cdot$s)
    }
    \label{Hp3_extrap}
\end{figure}

A reliable extraction of the ground state energy level in the $P=+1$ sector has been implemented using the variational quantum eigensolver (VQE) method \cite{Peruzzo2014} supplemented by classical Bayesian global optimization with Gaussian processes allowing for a minimal number of function calls to the quantum computer
(for other implementations, see Refs.~\cite{McClean2016,Santagati2018}). The structure of the
$P=+1$
Hamiltonian in Eq.~(\ref{eq:H1p1mats}) is that of a one-dimensional chain of $N = \tilde{\Lambda} +1$ sites with local chemical potentials $V_i$ and hopping amplitudes $t_{ij} = \sqrt{2}x$ for $|i-j|=1$ and $t_{01} = t_{10} = 2 x$.  The chemical potential varies from one site to the next site by $ 1  \pm 2 \mu$.  From this perspective, it is known that a series of local and controlled rotations can construct the resulting N-site, real eigenfunction.  VQE finds,
with linear error extrapolation in the noise parameter $r$,
the ground state energies of the ${\rm \bf{k}}=\bf{0}$ and $\tilde{\Lambda}=1,2,3$ spaces as $\langle H\rangle = -0.91(1)$~MeV, $-1.01(4)$~MeV, and $-1.01(2)$~MeV respectively (see Appendix~\ref{app:VQE}, \ref{app:CNOTerrors}, and \ref{app:chiralcondensate})\footnote{Example code snippets for calculation on IBM hardware and tables of data appearing in figures can be found in the supplemental material~\cite{Supplemental}}.
To manage inherent noise on the chip, we have performed computations with
a large number of measurement shots
(8192 shots for ibmqx2~\cite{IBMgates2} and ibmqx5~\cite{IBMgates5}).
For these variational calculations,
the systematic measurement errors have been corrected
via the readout-error mitigation strategy~\cite{kandala2017,PhysRevLett.120.210501}.
Further, a zero-noise extrapolation error mitigation technique inspired by Refs.~\cite{Li2017,Temme2017} has been implemented.
Examples of this zero-noise extrapolation technique are shown in
Fig.~\ref{Hp3_extrap}, where the noise parameter $r$ controls the accrual of systematic errors by inserting $r-1$ additional 2-qubit gates (CNOT$^2$) at every instance of a CNOT gate. In the limit of zero noise, this modifies CNOT simply by an identity.

For the results obtained on IBM quantum hardware, an estimate of the length of time
the quantum processing unit (QPU) spent executing instructions based upon IBM benchmarking is provided~\cite{IBMgates2,IBMgates5,Gambetta2012}.  This VQE calculation required 6.4 QPU-seconds and 2.4 CPU-seconds with a total run time of 4 hours.  Clearly, a majority of the time was spent in communications.
\section{Dynamical Properties}
\begin{figure}[!t]
	\includegraphics[width=\columnwidth]{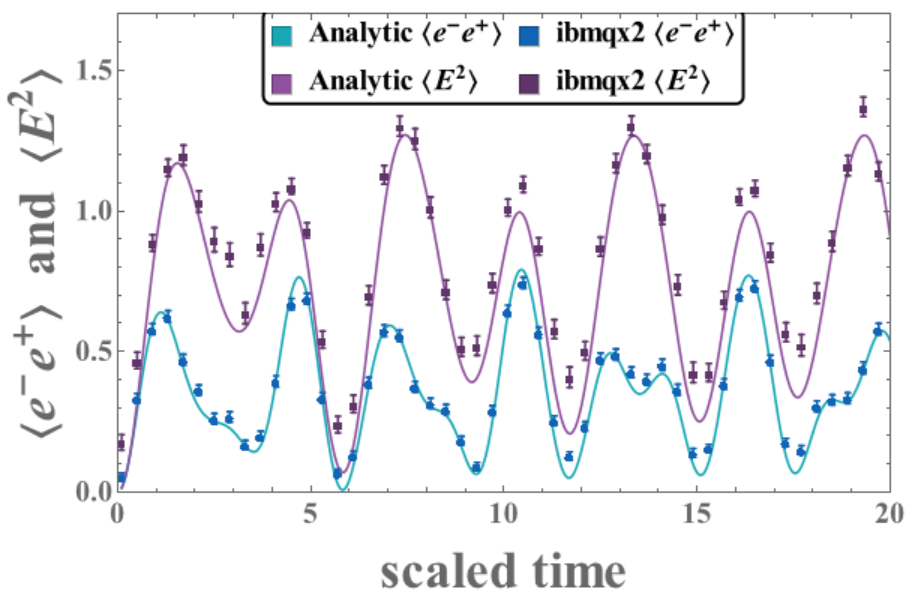}
	\caption{
	The probability of finding an $e^+e^-$ pair (blue, lower line) and the
	expectation value of the energy of the electric field (purple, upper line) in the two-spatial-site Schwinger model
	following time evolution with $U(\theta_i(t))$ from the initial empty state.
	The solid curves are exact results while the
	 the data points are quadratic extrapolations obtained with the ibmqx2 quantum computer using a circuit involving 3 CNOT gates~\cite{VidalDawson2004}.
	 (1000 IBM allocation units and $\sim12.3$~QPU$\cdot$s)
	}
	\label{fig:pairs1}
\end{figure}

Time evolving quantum systems is a key capability of quantum computers.
Working with the ${\bf k}={\bf 0}$ $P=+1$ sector, we  evolve the unoccupied state $ |\chi_1\rangle_{ {\bf k}={\bf 0},+} $ (see Fig.~\ref{fig:zp1} and Appendix~\ref{app:MomParCC})
forward in time with two techniques.
The first is through SU(4) parameterization of the evolution operator  and the second is using a Trotter discretization of time.
The former uses a classical computer to determine the 9 angles describing the
time evolution over an arbitrary time interval, which is induced by the symmetric SU(4) matrix $U(\theta_i(t)) = e^{-i {\rm H} t}$,
leading to the state
$|\chi\rangle_{ {\bf k}={\bf 0},+}(t)  = U(\theta_i; t) |\chi_1\rangle_{ {\bf k}={\bf 0},+}$
(see Appendix~\ref{app:SU4}).
The most general form of the symmetric SU(4) matrix through its Cartan decomposition is $U=K^T C K$ where
 $C = e^{-i \sigma_x\otimes\sigma_x \theta_7/2} e^{-i \sigma_y\otimes\sigma_y \theta_8/2} e^{-i \sigma_z\otimes\sigma_z \theta_9/2}$ is generated by the Cartan subalgebra
and  $K$ is a ${\rm SU(2)}\otimes {\rm SU(2)}$ transformation defined by the 6 angles, $\theta_{1,..6}$ \cite{Khaneja01cartandecomposition,Tilma2002}.
Fig.~\ref{fig:pairs1}
shows the ``zero-noise'' extrapolated
pair probability and expectation value of the energy in the electric field as a function of time
calculated on ibmqx2 with the Cartan subalgebra circuit of Ref.~\cite{VidalDawson2004}.

\begin{figure}[!t]
	\includegraphics[width=\columnwidth]{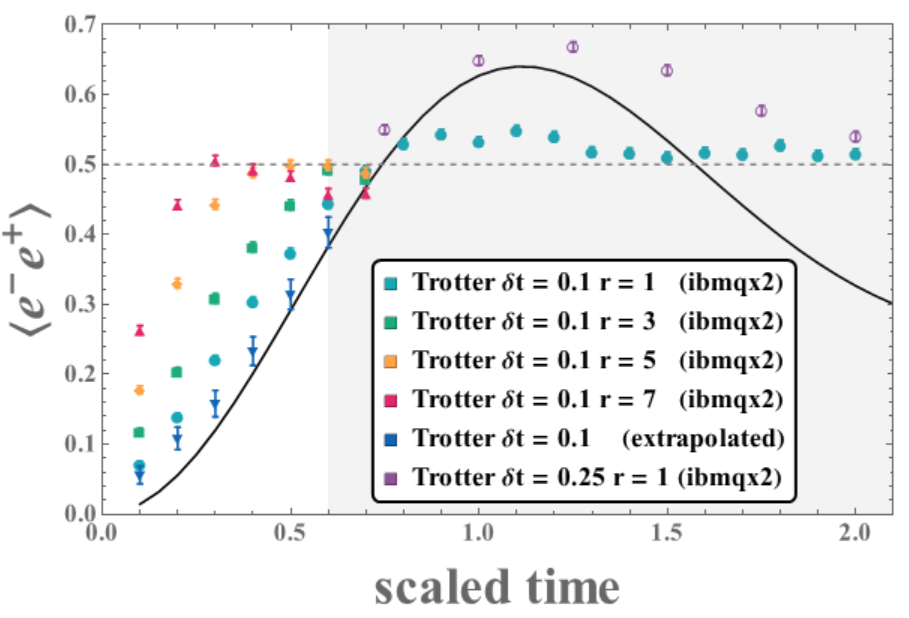}
	\caption{
	The probability of finding an $e^+e^-$ pair in the two-spatial-site Schwinger model from the initial empty state
	following time evolution with $U_T(t,\delta t)$. In the unshaded region, the blue points (triangle markers with visible error bars) are quadratic extrapolations to zero noise using the data above each point at increasing values of the noise parameter, $r$.
	(260 IBM allocation units and $\sim3.6$~QPU$\cdot$s)
	}
	\label{fig:pairs2}
\end{figure}

The time evolution of this system has also been studied using a Trotterized operator (see Appendix~\ref{app:trotter}). It is discretized such that $e^{-i {\rm H} t}\rightarrow U_T(t,\delta t) = \lim\limits_{N\rightarrow\infty} \left( \prod\limits_j e^{-i {\rm H}_j \delta t} \right)^N$, where $\delta t = \frac{t}{N}$ and the Hamiltonian decomposition ${\rm H} = \sum\limits_j{\rm H}_j$ (for the $\bf{k} = \bf{0}$ $P = +1$ $\tilde{\Lambda} = 3$ sector) is given by,
\begin{eqnarray}
{\rm H} & = &
\frac{x}{\sqrt{2}}\ \sigma_x\otimes\sigma_x
\ +\
\frac{x}{\sqrt{2}}\ \sigma_y\otimes\sigma_y
\ -\
\mu\ \sigma_z\otimes\sigma_z
\nonumber\\
&  &\ +  \
x\left(1 + \frac{1}{\sqrt{2}}\right) \ I \otimes\sigma_x
\ -\
\frac{1}{2}\ I \otimes\sigma_z
\nonumber\\
&  &\ -\
\left(1 + \mu\right)\ \sigma_z\otimes I
\ +
x\left(1 - \frac{1}{\sqrt{2}}\right) \ \sigma_z\otimes\sigma_x
\ \ .
\label{eq:hamdec}
\end{eqnarray}
We have optimized the sequence of operations
in a first-order Trotterization.  While Trotterization bypasses the classical resources needed in the previous time evolution implementation to solve for the 9 angles of a symmetric SU(4) matrix, its demand for long coherence times is not satisfied with the $T_2$ times available on current quantum hardware.  Using the reported gate specifications of ibmqx2 in terms of pulse sequences and their temporal extent, the $T_2$ coherence time of the device is reached after $\sim10$ time steps.  This can be seen in Fig.~\ref{fig:pairs2} where the Trotterized evolution with $\delta t = 0.1$ saturates to the classical probability of 0.5 after a small number of steps---quantum coherence has been lost.   This limitation in the number of coherent time steps encourages the use of larger values of $\delta t$ (top data in shaded region), trading accuracy of the Trotterization for coherence maintained further into the time evolution.  Even with this trade off, this method is currently unable to explore the low-energy structure of the dynamic fluctuations.
%
%\vspace*{3mm}

\section{Discussion and Outlook}
Our  work has identified key areas of future development needed to robustly explore  quantum field theories with (imperfect) universal quantum computers.  In order to explore more complex dynamics such as the scattering of hadrons or the time evolution of charge screening, a balance between the short-depth circuits of exact SU($2^n$) propagator evolution and the manageable classical resources required to Trotterize must be developed.  Regardless of the chosen method of time evolution, classical pre- and post-processing will continue to be invaluable for scientifically-relevant calculations on near-term quantum computers.
By enforcing Gauss's law,
momentum projecting states,
and imposing the discrete symmetry of parity,  the exponential growth of the Schwinger model Hilbert space has been softened sufficiently to achieve calculations on IBM’s superconducting quantum hardware.  This reduction has made possible the exploration of static and dynamic observables within the current and foreseeable experimental quantum computing landscape lacking quantum error correction and limited by coherence times and gate fidelities.  Requiring such a classical reduction in the process of building the physical, projected basis admittedly does not allow for advantage in the Hilbert space dimensionality accessible to the quantum vs classical computation.   However, the space of advantage is multidimensional.  By combining the strengths of the classical and quantum computers to respectively tame the Hilbert space and evolve it, the proposed heterogeneous framework profits in the exploration of time dependent, non-equilibrium, and finite density systems inaccessible to classical computations alone.

Our work represents one step toward solving QCD with NISQ era quantum computers
to address Grand Challenge problems in nuclear and high-energy physics.

%%%%%%%%%%%%%%%%%%%%%%%%%%%%%%%%%%%%%%%%%%%%%%%
\vspace*{2mm}

\begin{acknowledgments}
    We acknowledge use of the IBM Q experience for this work. The views expressed are those of the authors and do not reflect the official policy or position of IBM or the IBM Q experience team.
	We would like to thank Silas Beane, Aleksey Cherman,
	David Kaplan, John Preskill, Larry McLerran,
	Aidan Murran, Kenneth Roche, Alessandro Roggero,
	Jesse Stryker,
	Matthias Troyer and Nathan Weibe
	for many  important discussions and David Dean for helping to assemble the team.
	NK and MJS would like to thank the Institute for Quantum Information and Matter
	and Oak Ridge National Laboratory for kind hospitality during this work.
	MS and ES are grateful for funding through the Spanish MINECO/FEDER FIS2015-69983-P and Basque Government IT986-16.
	MJS and NK were supported  by DOE grant No.~DE-FG02-00ER41132. NK was supported in part by the Seattle Chapter
    of the Achievement Rewards for College Scientists
    (ARCS) foundation. This work is supported by the U.S. Department of Energy, Office of Science, Office of Advanced Scientific Computing Research (ASCR) quantum algorithms and testbed programs, under field work proposal numbers ERKJ333 and ERKJ335. This work was performed in part at Oak Ridge National Laboratory, operated by UT-Battelle for the U.S. Department of Energy under Contract No. DEAC05-00OR22725.
	
\end{acknowledgments}

\onecolumngrid
\begin{appendix}

\section{Momentum, Parity and Charge Conjugation}
\label{app:MomParCC}
In addition to the local U(1) gauge symmetry associated with the electromagnetic interaction, the Schwinger model~\cite{Schwinger:1962tp,Coleman:1975pw} respects a number of discrete symmetries.  Of particular interest and importance in this work are lattice representations of {\it parity} and
{\it charge conjugation}, denoted by operators $\hat P_a$ and $\hat C_i$ respectively (we will discuss the subscripts subsequently).
These operators commute with the Hamiltonian, and as such the eigenstates of systems
can be classified with respect to their transformations under $\hat C_i$ and $\hat P_a$, and (trivially) the combined operation of $\hat C_i\hat P_a$.
In the staggered (Kogut-Susskind) discretization~\cite{KogutSusskind1975,Banks:1975gq,Casher:1973uf}, the transformation properties of the fermion field operators and the electromagnetic field are well known under
$\hat C$ and $\hat P$ \cite{Hauke2013,VandanDoel1983,Golterman1984}, and we do not repeat them here.
However, we will discuss their implications for the systems we are examining in a little more detail.
The operation of $\hat C$ transforms  particles into antiparticles and vice versa, and the direction of the electric field  reverses as a result.
In order to maintain a physical representation in the Jordan-Wigner formulation~\cite{Jordan:1928wi} of staggered fermions~\cite{Banks:1975gq}, an additional directional shift by one lattice site (1/2 a spatial site) is necessary, with the direction being convention dependent.
The $\hat P_a$ transformation corresponds to
reflecting the system through axes, ``a'', that preserves the structure of the Wigner-Jordan representation of the fermion fields.

\begin{figure}[!ht]
\includegraphics[width=0.6\textwidth]{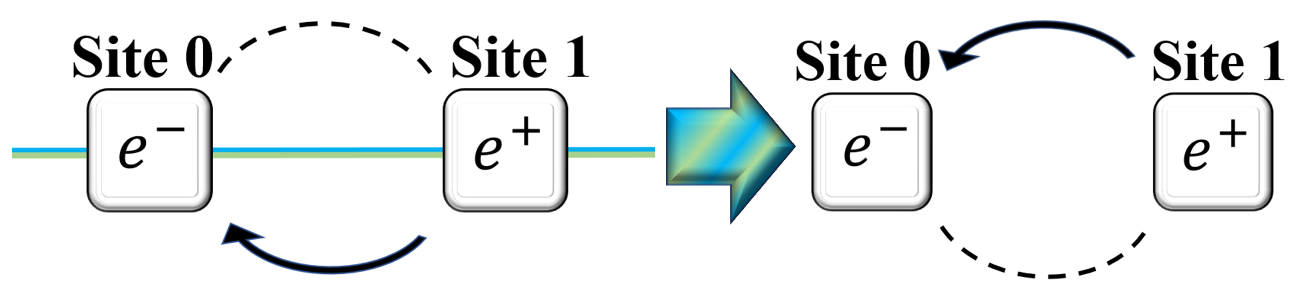}
\vskip 0.2in
\includegraphics[width=0.6\textwidth]{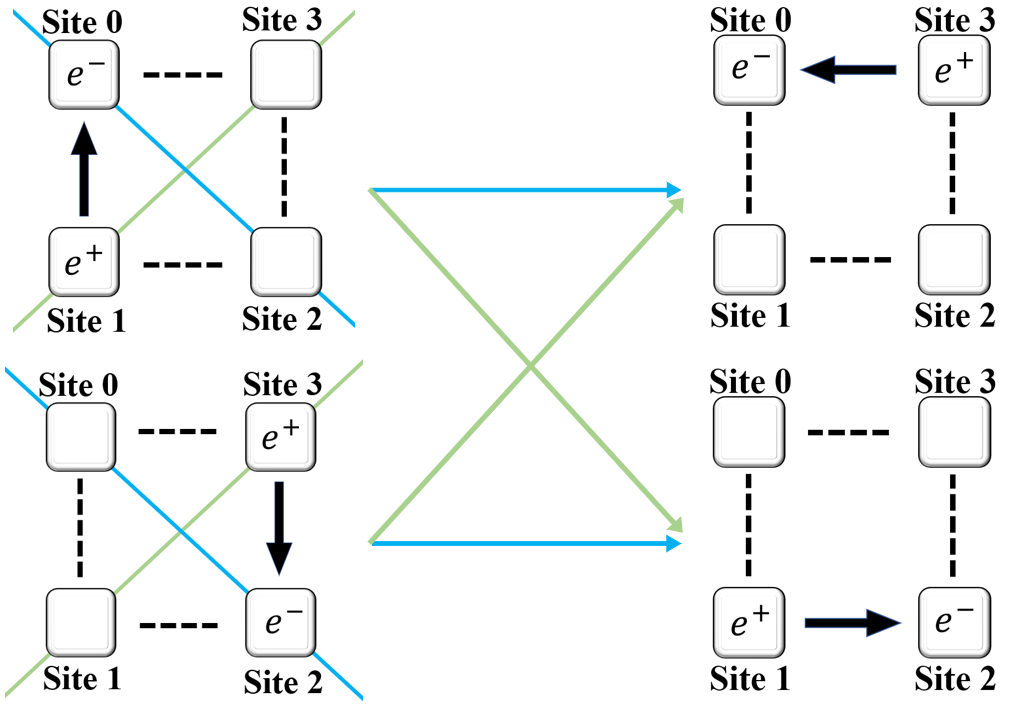}
\caption{
Examples of the action of the parity operators defined by the ``electron'' axes (blue lines, horizontal arrows and site 0-2 symmetry axis) and ``positron'' axes (green lines, diagonal arrows and site 1-3 symmetry axis).
An $e^-$ or an $e^+$ in one of the squares at a site indicates that the particle is present.  An arrow indicates an electric flux link aligned with the arrow, while a dashed link corresponds to the absence of an electric flux link.
In the $0+1$ example (upper panel) the only symmetry axis passes through both an electron and positron.
In the $1+1$ example (lower panel) there are two symmetry axes, one through the electron sites, and one through the positron sites.
}
\label{fig:parity}
\end{figure}

We begin by considering the $1+1$ system with two spatial sites. There are 13 physical states that satisfy Gauss's law in the charge $Q=0$ sector:
\begin{equation}\label{eq:1p1states}
  \begin{split}
|\phi_1\rangle & =  |\cdot\cdot\cdot\cdot\rangle |0000\rangle
\\
|\phi_2\rangle & =  |\cdot\cdot\cdot\cdot\rangle |1111\rangle
\\
|\phi_3\rangle & =  |\cdot\cdot\cdot\cdot\rangle |-1-1-1-1\rangle
\\
|\phi_4\rangle & =  |e^-e^+\cdot\cdot\rangle |-1000\rangle
\\
|\phi_5\rangle & =  |\cdot\cdot e^-e^+\rangle |00-10\rangle
\\
|\phi_6\rangle & =  |e^-e^+\cdot\cdot\rangle |0111\rangle
\\
|\phi_7\rangle & =  | \cdot\cdot e^-e^+\rangle |1101\rangle
  \end{split}
  \qquad
  \begin{split}
 |\phi_8\rangle & =  |e^-e^+ e^-e^+ \rangle |-10-10\rangle
\\
 |\phi_9\rangle & =  |e^-e^+ e^-e^+ \rangle |0101\rangle
\\
 |\phi_{10}\rangle & =  |e^- \cdot \cdot e^+ \rangle |-1-1-10\rangle
 \\
 |\phi_{11}\rangle & =  |e^- \cdot \cdot e^+ \rangle |0001\rangle
\\
 |\phi_{12}\rangle & =  | \cdot e^+  e^- \cdot  \rangle |0100 \rangle
\\
 |\phi_{13}\rangle & =  | \cdot e^+  e^- \cdot  \rangle |-10-1-1 \rangle
 \ \ ,
  \end{split}
\end{equation}
where a ``$\cdot$'' denotes an unoccupied site.
With periodic boundary conditions (PBCs), this system should be
considered as a square with the fermion sites at each corner.
For this system, there are two reflection axes that preserve the structure of the discretization, a reflection in the diagonal line defined by the electrons, and a reflection in the diagonal line defined by the positrons.
These parity transformations are shown in the lower panel of
Fig.~\ref{fig:parity}.

It is informative to consider the action of the charge conjugation operators
$\hat C_\pm$.
Along with the interchange of $e^+\leftrightarrow e^-$, there is a shift by half a spatial lattice site in either direction that is required to preserve the qubit structure.
For example,
\begin{eqnarray}
\hat C_+ |e^-e^+\cdot\cdot\rangle |-1000\rangle
& = &
|\cdot e^+ e^-\cdot\rangle |0100\rangle
\nonumber\\
\hat C_- |e^-e^+\cdot\cdot\rangle |-1000\rangle
& = &
|e^-\cdot\cdot e^+\rangle |0001\rangle
\end{eqnarray}

As the eigenstates of the Hamiltonian naturally arrange themselves into sectors of definite momentum, ${\bf k}$,
constrained to satisfy  ${\bf k} = \pi n$ with $n=0,\pm 1$
for the two spatial site system, it is convenient to first define states of good momentum.  To construct the states with
${\bf k}={\bf 0}$,
each state in Eq.~(\ref{eq:1p1states}) is rotated by two fermion sites (one spatial site) and added to the original state, with the sum appropriately renormalized.
This leads to a system involving 9 states:
\begin{equation}
  \begin{split}
|\psi_1\rangle_{{\bf k}={\bf 0}} & =  |\phi_1\rangle
\\
|\psi_2\rangle_{{\bf k}={\bf 0}} & =  |\phi_2\rangle
\\
|\psi_3\rangle_{{\bf k}={\bf 0}} & =  |\phi_3\rangle
\\
|\psi_4\rangle_{{\bf k}={\bf 0}} & =  {1\over\sqrt{2}}\ \left[\ |\phi_4\rangle + |\phi_5\rangle  \ \right]
\\
|\psi_5\rangle_{{\bf k}={\bf 0}} & =  {1\over\sqrt{2}}\ \left[\ |\phi_6\rangle + |\phi_7\rangle \  \right]
  \end{split}
  \qquad
  \begin{split}
|\psi_6\rangle_{{\bf k}={\bf 0}} & =  |\phi_8\rangle
\\
|\psi_7\rangle_{{\bf k}={\bf 0}} & =  |\phi_9\rangle
\\
|\psi_8\rangle_{{\bf k}={\bf 0}} & =  {1\over\sqrt{2}}\ \left[\ |\phi_{10}\rangle + |\phi_{13}\rangle \  \right]
\\
|\psi_9\rangle_{{\bf k}={\bf 0}} & =  {1\over\sqrt{2}}\ \left[\ |\phi_{11}\rangle + |\phi_{12}\rangle \ \right]
\ \ .
  \end{split}
  \label{eq:1p1k0states}
\end{equation}
Applying the two distinct parity operators to the momentum projected states results in the same states, and therefore, only one of the parity operators need be considered. In the zero momentum sector, the parity operator maps the states back into the same sector and has the same action as the charge conjugation operator,
and therefore $\hat C\hat P = +1$ for all states in this sector.
Forming states of good parity, by forming combinations of these 9 states with their parity transformed partner with a relative sign of $\pm 1$, leads to two sectors,
a 5-dimensional even parity sector, and a 4-dimensional odd-parity sector:
\begin{equation}
  \begin{split}
|\chi_1\rangle_{{\bf k}={\bf 0},+} & =  |\psi_1\rangle
\\
|\chi_2\rangle_{{\bf k}={\bf 0},+}  & =  {1\over\sqrt{2}}\ \left[\ |\psi_{4}\rangle + |\psi_{9}\rangle \  \right]
\\
|\chi_3\rangle_{{\bf k}={\bf 0},+}  & =  {1\over\sqrt{2}}\ \left[\ |\psi_{6}\rangle + |\psi_{7}\rangle \  \right]
  \end{split}
  \qquad
  \begin{split}
|\chi_4\rangle_{{\bf k}={\bf 0},+}  & =  {1\over\sqrt{2}}\ \left[\ |\psi_{5}\rangle + |\psi_{8}\rangle \  \right]
\\
|\chi_5\rangle_{{\bf k}={\bf 0},+}  & =  {1\over\sqrt{2}}\ \left[\ |\psi_{2}\rangle + |\psi_{3}\rangle \  \right]
\ \ ,
\\
  \end{split}
  \label{eq:1p1k0Estates}
\end{equation}
and
\begin{equation}
  \begin{split}
|\chi_1\rangle_{{\bf k}={\bf 0},-}  & =  {1\over\sqrt{2}}\ \left[\ |\psi_{4}\rangle - |\psi_{9}\rangle \  \right]
\\
|\chi_2\rangle_{{\bf k}={\bf 0},-}  & =  {1\over\sqrt{2}}\ \left[\ |\psi_{6}\rangle - |\psi_{7}\rangle \  \right]
  \end{split}
  \qquad
  \begin{split}
|\chi_3\rangle_{{\bf k}={\bf 0},-}  & =  {1\over\sqrt{2}}\ \left[\ |\psi_{5}\rangle - |\psi_{8}\rangle \  \right]
\\
|\chi_4\rangle_{{\bf k}={\bf 0},-}  & =  {1\over\sqrt{2}}\ \left[\ |\psi_{2}\rangle - |\psi_{3}\rangle \  \right]
\ \ ,
  \end{split}
  \label{eq:1p1k0Ostates}
\end{equation}

For the
${\bf k}=\pm {\bf 1}$ states, the process is analogous to the zero-momentum sector, except that the translated state is multiplied by $-1$ before being added to the original state,
\begin{equation}
  \begin{split}
|\psi_1\rangle_{|{\bf k}| = 1} & =  {1\over\sqrt{2}}\ \left[\ |\phi_4\rangle - |\phi_5\rangle  \ \right]
\\
|\psi_2\rangle_{|{\bf k}| = 1} & =  {1\over\sqrt{2}}\ \left[\ |\phi_6\rangle - |\phi_7\rangle \  \right]
  \end{split}
  \qquad
  \begin{split}
|\psi_3\rangle_{|{\bf k}| = 1} & =  {1\over\sqrt{2}}\ \left[\ |\phi_{10}\rangle - |\phi_{13}\rangle \  \right]
\\
|\psi_4\rangle_{|{\bf k}| = 1} & =  {1\over\sqrt{2}}\ \left[\ |\phi_{11}\rangle - |\phi_{12}\rangle \ \right]
\ \ .
  \end{split}
  \label{eq:1p1kpm1states}
\end{equation}

In the case of the $0+1$ system, with only one spatial site, the only symmetry axis about which reflections can be performed that leave the qubit structure intact is through the axis defined by the qubits themselves. This leads to reflections between the electric flux links only.
The 5 states in this system that satisfy Gauss's law are
\begin{equation}
  \begin{split}
|\phi_1\rangle & =  |\cdot\cdot\rangle |-1-1\rangle
\\
|\phi_2\rangle & =  |\cdot\cdot\rangle |00\rangle
\\
|\phi_3\rangle & =  |\cdot\cdot\rangle |11\rangle
  \end{split}
  \qquad
  \begin{split}
|\phi_4\rangle & =  | e^- e^+ \rangle |01\rangle
\\
|\phi_5\rangle & =  | e^- e^+ \rangle |-10\rangle
 \ \ ,
  \end{split}
  \label{eq:0p1states}
\end{equation}
which, without the possibility of momentum projection, decompose into the two parity sectors.
The even-parity sector is composed of 3 states, while the odd-parity sector is composed of 2 states:
\begin{equation}
  \begin{split}
|\psi_1\rangle_+ & =  |\cdot\cdot\rangle |00\rangle
\\
|\psi_2\rangle_+ & =  | e^- e^+ \rangle  {1\over\sqrt{2}} \left[\ |01\rangle + |-10\rangle \ \right]
\\
|\psi_3\rangle_+ & =  |\cdot\cdot\rangle  {1\over\sqrt{2}} \left[\ |11\rangle + |-1-1\rangle \ \right]
  \end{split}
  \qquad
  \begin{split}
|\psi_1\rangle_- & =  | e^- e^+ \rangle  {1\over\sqrt{2}} \left[\ |01\rangle - |-10\rangle \ \right]
\\
|\psi_2\rangle_- & =  |\cdot\cdot\rangle  {1\over\sqrt{2}} \left[\ |11\rangle - |-1-1\rangle \ \right]
 \ \ ,
  \end{split}
  \label{eq:0p1EOstates}
\end{equation}

It is interesting to note that the parity transformations we have discussed in this section extend to systems with more spatial sites, subject to the constraint that the number of spatial sites is a multiple of two. This makes it natural to extend our studies to systems with $N_{\rm sites} = 4,6,8,...$.

%%%%%%%%%%%%%%%%%%%%%%%%%%%%%%%%%
\section{Exact two-site Schwinger Model Spectra}
\label{app:exact2site}

\begin{figure}[!ht]
\includegraphics[width=0.5\textwidth]{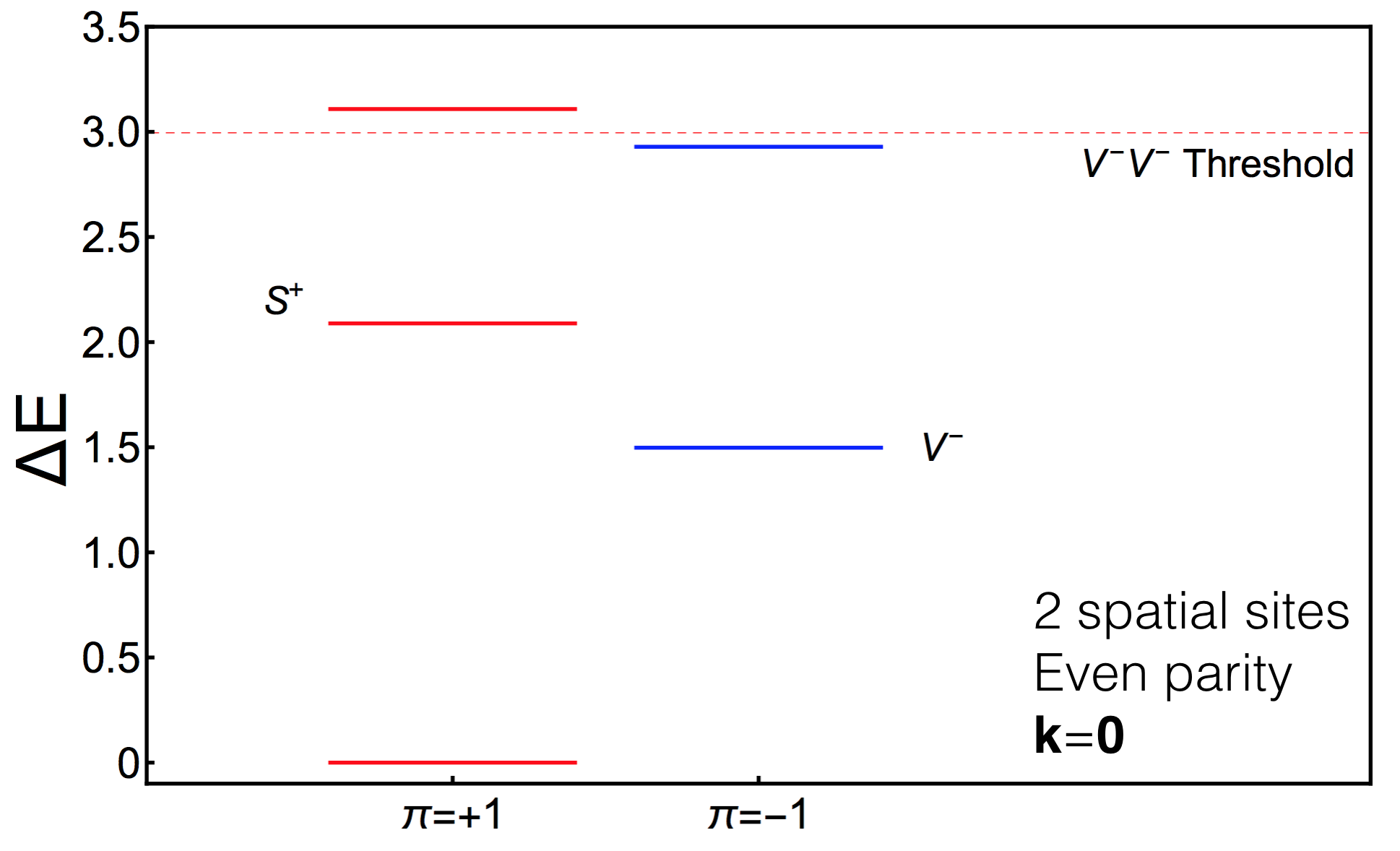}
\caption{
The low-lying spectrum of the $1+1$ Schwinger model
discretized onto a lattice with two spatial sites
with couplings $\mu=0.1$, $x=0.6$, and projected to zero momentum.
The shown shifted $P=+1$ energy eigenvalues are
$0$, $2.089$ and $3.108$
and the $P=-1$ energy eigenvalues are
$1.497$ and $2.927$.
}
\label{fig:N8SSpec}
\end{figure}

The spectrum of the Schwinger model is rich.
As our calculations are performed at a single lattice spacing without a continuum extrapolation, and with one and two spatial sites without an infinite volume extrapolation, it is helpful to discuss what is to be expected from them.
The spectrum of the  Schwinger model
discretized onto a lattice with two spatial sites
with couplings $\mu=0.1$, $x=0.6$
and cut off $\tilde\Lambda=10$,
is shown in Fig.~\ref{fig:N8SSpec}.
The ground state energy has been defined (shifted)  to be zero, but on an absolute scale is $E_0=-1.011\ 810$, corresponding to an energy density of $\epsilon_0=-0.505\ 905$.  Further, there is a chiral condensate, $\langle\overline{\psi}\psi\rangle = -0.322\ 324$.
The first excited state is odd-parity, defined to be the lightest vector meson, $V^-$, (the massive photon), and the second excited state is even parity, defined to be the scalar meson, $S^+$.
The next even-parity excited state in the spectrum is just above the $V^-V^-$ threshold, and corresponds to two vector mesons with a repulsive interaction between them.  The splitting from the threshold is a finite volume effect and vanishes as the volume of space tends to infinity.
It is analogues of
this energy splitting that are used successfully in lattice QCD calculations in Euclidean space,
in conjunction with quantum field theory quantization conditions~\cite{Luscher:1986pf,Luscher:1990ux},
to determine scattering phase shifts and mixing parameters between the strongly interacting hadrons of QCD (for recent examples of such calculations, see Refs.~\cite{Briceno:2017qmb,Wagman:2017tmp}).
In addition, higher in the spectrum of larger systems, there is a state that corresponds to a very loosely bound three-body system.

The volume scaling of vacuum properties demonstrate their expected exponential convergence, as can be seen from Table~\ref{tab:vacuum}.
While the vacuum energy is an extensive quantity, the energy density rapidly converges
to a constant value, and is within $\sim 1\%$ of its infinite volume value with
two spatial sites for the parameters we have chosen.
\begin{table}[!ht]
		\begin{tabular}{c|cccc}
\hline
$\#$ of Spatial Sites  & 2 & 4 & 6 & 8 \\ \hline
$E_{\rm vac}$  & -1.011 810 \  &  -2.019 632 \  & -3.029 438 \  & -4.039 251    \\
$\varepsilon_{\rm vac}$ & -0.505 905 \  &  -0.504 908 \  & -0.504 906 \  & -0.504 906   \\
$\langle \overline{\psi}\psi\rangle$ & -0.322 324  \  & -0.324 713  \    &  -0.324 722 \   & -0.324 722 \\
$\langle E^2 \rangle$ & 0.089 457 \    & 0.088 044  \   & 0.088 039  \   &  0.088 039   \\
\hline
		\end{tabular}
\caption{
\label{tab:vacuum}
Ground state properties of the $1+1$ Schwinger model.
The vacuum energy, vacuum energy density, chiral condensate and total energy in the electric field,
for $\mu=0.1$, $x=0.6$ and a cut off of $\tilde{\Lambda}=10$  in the electric field,
for a selection of the number of spatial sites.
}
\end{table}
%

%%%%%%%%%%%%%%%%%%%%%%%%%%%%%%%%%%%%%%%%%%%%%%%
\section{SU(4) Transformations for 2-qubits}
\label{app:SU4}
Elements of the SU(N) Lie-group can be obtained by exponentiating its ${\rm N}^2-1$ generators, each multiplied by a real angle.
With four states in the fundamental representation,
the unitary rotations of two qubits are described by SU(4),  requiring 15 angles to be specified.  A succinct parameterization
of these transformations is given in the Pauli basis, as presented by Khaneja and Glaser~\cite{Khaneja01cartandecomposition},
and compactly written as
\begin{equation}
  U = K_2 \ e^{-i \left(\alpha_1\sigma_x\otimes \sigma_x + \alpha_2\sigma_y \otimes \sigma_y +\alpha_3 \sigma_z \otimes \sigma_z\right)} \ K_1
  \ =\
  K_2 \ C \ K_1
\end{equation}
with $K_{1,2} \in \text{SU(2)} \otimes \text{SU(2)}$, where the SU(2)'s act on the individual qubits,
and $C$ denotes transformations associated with the Cartan sub-algebra.
For time evolution, the symmetric forms of the Hamiltonian matrices we are working with lead to only symmetric SU(4) transformations,
 while for variational state preparation, relative phases between states in the eigenbasis may be removed.
 Enforcing symmetry on an SU(4) transformation matrix reduces the number of angles  from 15 to 9 (through the 6 constraints),
 and eliminating the relative phases between the states further reduces the number of angles from 9 to 6.

%%%%%
\par
\begin{figure}[h]
  \includegraphics[width=0.6\textwidth]{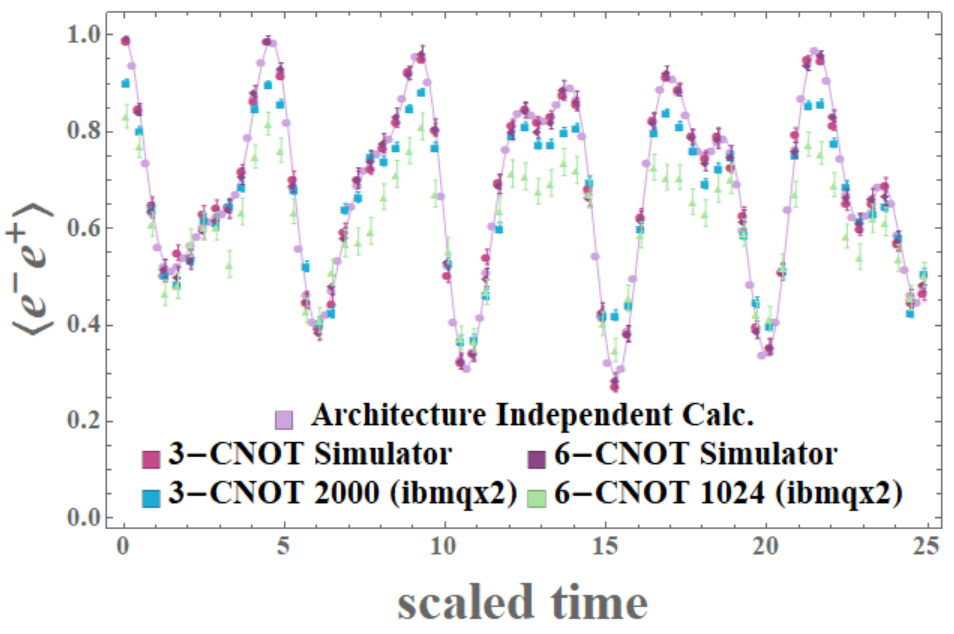}
  \caption{
  The probability of having one $e^+e^-$ pair in the
  $1+1$, odd-parity system at some time after starting in the lowest-energy basis state containing one $e^+e^-$ pair.
  The state is evolved forward by a single application of the exact propagator described in this section.
  These probabilities were determined on both the IBM simulator(s) and quantum hardware, ibmqx2.  Two different circuits were used to implement the transformations from the Cartan subalgebra, one with 3 CNOT gates (blue squares) and one with 6 CNOT gates (green triangles).  (504 IBM allocation units were used for the $\sim0.7$~QPU$\cdot$s needed to generate this data set.)
  }
  \label{fig:1pairfluct}
\end{figure}

The symmetric SU(4) transformations may be parameterized by relating the angles of $K_2$ to those of $K_1$,
\[U^T = \left(K_2\  C\  K_1\right)^T = K_1^T\  C\  K_2^T \ \ \Rightarrow \ \ K_2 = K_1^T\]
Using the standard ZYZ (Euler angles) parameterization for each $SU(2)$,
\begin{align}
K_1 &=  e^{-i \frac{\theta_6}{2} \mathbb{I}\otimes\sigma_z} e^{-i \frac{\theta_5}{2} \mathbb{I}\otimes\sigma_y} e^{-i \frac{\theta_4}{2} \mathbb{I}\otimes\sigma_z}e^{-i \frac{\theta_3}{2} \sigma_z\otimes\mathbb{I}}e^{-i \frac{\theta_2}{2} \sigma_y\otimes\mathbb{I}}e^{-i \frac{\theta_1}{2} \sigma_z\otimes\mathbb{I}} \\
&= \begin{gathered}
\Qcircuit @C=0.8em @R=-0.5cm @! {
  & \gate{e^{-i \frac{\theta_1}{2} \sigma_z}} & \gate{e^{-i \frac{\theta_2}{2} \sigma_y}} & \gate{e^{-i \frac{\theta_3}{2} \sigma_z}} & \qw \\
  &\gate{e^{-i \frac{\theta_4}{2} \sigma_z}} & \gate{e^{-i \frac{\theta_5}{2} \sigma_y}} & \gate{e^{-i \frac{\theta_6}{2} \sigma_z}} & \qw
}
\end{gathered}
\end{align}
\begin{align}
K_2 = K_1^T &= e^{-i \frac{\theta_4}{2} \mathbb{I}\otimes\sigma_z} e^{i \frac{\theta_5}{2} \mathbb{I}\otimes\sigma_y} e^{-i \frac{\theta_6}{2} \mathbb{I}\otimes\sigma_z}e^{-i \frac{\theta_1}{2} \sigma_z\otimes\mathbb{I}}e^{i \frac{\theta_2}{2} \sigma_y\otimes\mathbb{I}}e^{-i \frac{\theta_3}{2} \sigma_z\otimes\mathbb{I}} \\
&= \begin{gathered}
\Qcircuit @C=0.8em @R=-0.5cm @! {
  & \gate{e^{-i \frac{\theta_3}{2} \sigma_z}} & \gate{e^{i \frac{\theta_2}{2} \sigma_y}} & \gate{e^{-i \frac{\theta_1}{2} \sigma_z}} & \qw \\
  &\gate{e^{-i \frac{\theta_6}{2} \sigma_z}} & \gate{e^{i \frac{\theta_5}{2} \sigma_y}} & \gate{e^{-i \frac{\theta_4}{2} \sigma_z}} & \qw
}
\end{gathered}
\end{align}
and an arbitrary symmetric 2-qubit transformation, defined by 9 angles,
may be parameterized as:
\begin{multline}
  U_p = e^{-i \frac{\theta_4}{2} \mathbb{I}\otimes\sigma_z} e^{i \frac{\theta_5}{2} \mathbb{I}\otimes\sigma_y} e^{-i \frac{\theta_6}{2} \mathbb{I}\otimes\sigma_z}e^{-i \frac{\theta_1}{2} \sigma_z\otimes\mathbb{I}}e^{i \frac{\theta_2}{2} \sigma_y\otimes\mathbb{I}}e^{-i \frac{\theta_3}{2} \sigma_z\otimes\mathbb{I}}
  e^{-i \frac{\theta_9}{2} \sigma_z \otimes \sigma_z}e^{-i \frac{\theta_8}{2} \sigma_y \otimes \sigma_y} e^{-i \frac{\theta_7}{2} \sigma_x \otimes \sigma_x} \\ e^{-i \frac{\theta_6}{2} \mathbb{I}\otimes\sigma_z} e^{-i \frac{\theta_5}{2} \mathbb{I}\otimes\sigma_y} e^{-i \frac{\theta_4}{2} \mathbb{I}\otimes\sigma_z}e^{-i \frac{\theta_3}{2} \sigma_z\otimes\mathbb{I}}e^{-i \frac{\theta_2}{2} \sigma_y\otimes\mathbb{I}}e^{-i \frac{\theta_1}{2} \sigma_z\otimes\mathbb{I}}
  \ .
\end{multline}
If a system is initially prepared in a state in the computational, z-axis basis,
the first $\sigma_z$ rotation on each qubit simply induces an overall phase in the wavefunction, and hence can be dropped.
\begin{multline}
  U_p = e^{-i \frac{\theta_4}{2} \mathbb{I}\otimes\sigma_z} e^{i \frac{\theta_5}{2} \mathbb{I}\otimes\sigma_y} e^{-i \frac{\theta_6}{2} \mathbb{I}\otimes\sigma_z}e^{-i \frac{\theta_1}{2} \sigma_z\otimes\mathbb{I}}e^{i \frac{\theta_2}{2} \sigma_y\otimes\mathbb{I}}e^{-i \frac{\theta_3}{2} \sigma_z\otimes\mathbb{I}}
  e^{-i \frac{\theta_9}{2} \sigma_z \otimes \sigma_z}e^{-i \frac{\theta_8}{2} \sigma_y \otimes \sigma_y} e^{-i \frac{\theta_7}{2} \sigma_x \otimes \sigma_x} \\ e^{-i \frac{\theta_6}{2} \mathbb{I}\otimes\sigma_z} e^{-i \frac{\theta_5}{2} \mathbb{I}\otimes\sigma_y} e^{-i \frac{\theta_3}{2} \sigma_z\otimes\mathbb{I}}e^{-i \frac{\theta_2}{2} \sigma_y\otimes\mathbb{I}}
  \ .
\end{multline}
In order to implement the rotations of the Cartan subalgebra, two options were explored: 6 CNOTs with the textbook implementation of rotations~\cite{NielsenChuang} for each generator
or 3 CNOTs as implemented in Vidal and Dawson~\cite{VidalDawson2004}
and by Coffey {\it et. al.}~\cite{Coffey2008},
\begin{multline}
  e^{-\frac{i}{2}\left(\theta_7 \sigma_x \otimes \sigma_x + \theta_8 \sigma_y \otimes \sigma_y + \theta_9 \sigma_z \otimes \sigma_z\right)} =  \\ \begin{gathered}
  \Qcircuit @C=-1.1em @R=-0.5cm @! {
  &  \gate{H} & \ctrl{1} & \qw & \ctrl{1} & \gate{H} & \gate{S^\dagger} & \gate{H} & \ctrl{1} & \qw & \ctrl{1} & \gate{H} & \gate{S} & \ctrl{1} & \qw & \ctrl{1} & \qw\\
  & \gate{H} & \targ & \gate{e^{-i \frac{\theta_7}{2} \sigma_z }} & \targ & \gate{H} & \gate{S^\dagger} & \gate{H} & \targ & \gate{e^{-i \frac{\theta_8}{2} \sigma_z }} & \targ & \gate{H} & \gate{S} & \targ & \gate{e^{-i \frac{\theta_9}{2} \sigma_z }} & \targ & \qw
}
  \end{gathered}
\end{multline}
\begin{equation}
      e^{-\frac{i}{2}\left(\theta_7 \sigma_x \otimes \sigma_x + \theta_8 \sigma_y \otimes \sigma_y + \theta_9 \sigma_z \otimes \sigma_z\right)} =  \begin{gathered}
\Qcircuit @C=-0.7em @R=-0.5cm @! {
  &  \ctrl{1} & \gate{e^{-i\frac{\theta_7}{2} \sigma_x}}  & \gate{H} & \ctrl{1} & \gate{S} & \gate{H} & \ctrl{1} & \gate{e^{i \frac{\pi}{4} \sigma_x}} & \qw\\
  & \targ & \gate{e^{-i \frac{\theta_9}{2} \sigma_z }}  & \qw & \targ & \gate{e^{i \frac{\theta_8}{2} \sigma_z}} & \qw & \targ & \gate{e^{-i \frac{\pi}{4} \sigma_x}} & \qw
}
\end{gathered}
\end{equation}
Though technically equivalent and returning consistent results in simulations, the above two circuits have  different signatures of systematic errors when executed on  quantum computing hardware.  The difference can be seen in Figure~\ref{fig:1pairfluct} where the systematic errors at high probabilities are exacerbated when using the 6-CNOT circuit (which also includes a number of additional operations).

%%%%%%%%%%%%%%%%%%%%%%%%%%%%%%%%%%%%%%%%
\section{Trotterization}
\label{app:trotter}
In the previous section, we determined the exact propagator (in terms of 9 angles) that evolves an arbitrary 2-qubit state forward over a macroscopic time interval.
While the theoretical accuracy and gate requirements of simulating dynamical quantum systems defined on $n$ qubits,
with exact propagators as symmetric matrices in SU($2^n$), can be determined, the associated dimensionality of the parameter space of the angles is $2^{n-1}\left(2^n+1\right)-1$.
This growth in the number of angles that need to be determined with classical computing resources
appears to be unsustainable for classical optimization when looking forward to large quantum computers.
For this well-known reason, Trotterizing the time evolution operator appears to be a necessary technique for exploring quantum systems.

In first-order Trotterization, the time evolution operator is approximated by breaking apart the exponential and suppressing the resulting commutators in powers of $H\delta t = H\frac{t}{N_{\rm Trot.}}$ where $t$ is the total time propagated and $N_{\rm Trot.}$ is the number of time steps into which the propagator is divided,
\begin{equation}
  e^{-i H t} = e^{-i \sum\limits_j H_j t} =\lim_{N_{\rm Trot.} \rightarrow \infty} \left(\prod\limits_j e^{-i H_j \delta t
 }\right)^{N_{\rm Trot.}}
 \ \ \ .
 \label{eq:HTrott}
\end{equation}
While large resources and long coherence times would
allow structure from terms sub-leading in $H\delta t$ to be made inconsequential, the results of Trotterization on the 0+1 and 1+1 dimensional Schwinger model
indicate that we are not yet able to accomplish this with IBM quantum computing hardware.
In near-term quantum computations, care must be given to balance the theoretical errors built into the Trotterization of the evolution operator with the gate fidelities and with the coherence times of the hardware.  In a recent publication~\cite{Hadfield2018}, an idea for multi-step Trotterization to focus resources on physically-dominant terms in the Hamiltonian has been proposed and analyzed for its improved scaling properties of quantum simulation.  Such strategies to optimally utilize simulation resources will be important for optimizing scientific output from any quantum hardware.
\begin{figure}[!ht]
  \includegraphics[width = 0.40\textwidth]{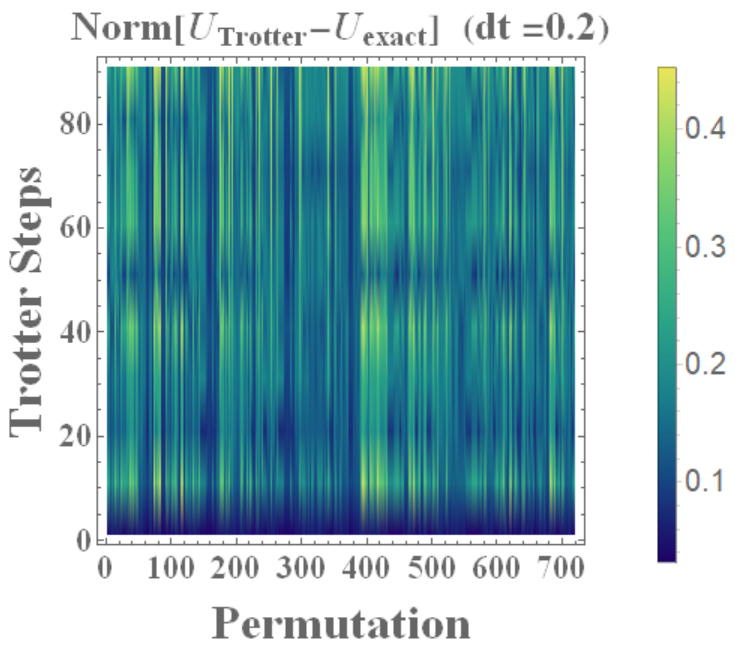}
  \includegraphics[width = 0.50\textwidth]{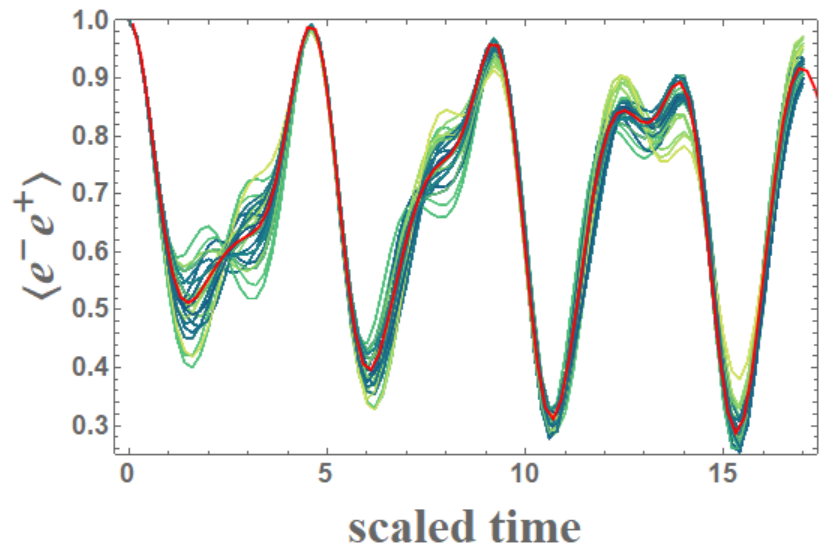}
  \caption{
  The left panel shows the normed difference between the exact propagator and the Trotterized propagator with a step size of $\delta t = 0.2$ for different permutation orders of the Hamiltonian terms in Eq.~(\ref{eq:HTrott}).
  The right panel shows the $e^+e^-$ pair probability as a function of time for a selection of orderings of the Trotterized propagator.
  }
  \label{fig:Trotts}
\end{figure}
By classical simulation, we performed a rudimentary Trotterization optimization by sampling over orderings of the component contributions to the Trotterized propagator in Eq.~(\ref{eq:HTrott}) for the $4\times 4$ Hamiltonian matrix describing the ${\bf k}=\bf{0}$ and $P=+1$ sector of the $1+1$ Schwinger model.
The results of these calculations are shown in Fig.~\ref{fig:Trotts}.

%%%%%%%%%%%%%%%%%%%%%%%%%%%%%%%%%

%%%%%%%%%%%%%%%%%%%%%%%%%%%%%%
\section{Variational Calculations of  Energy Eigenvalues}
\label{app:VQE}
To provide an example of our variational calculations of the energy eigenvalues, we use the $1+1$  Schwinger model restricted to the $P=+1$, ${\bf k}={\bf 0}$ sector.
By eliminating the state with the largest energy in the electric field,
the $5\times 5$ Hamiltonian matrix is truncated to a $4\times 4$ matrix, which can be studied with two qubits.
The Hamiltonian in this truncated space is,
\begin{eqnarray}
 H_{{\bf k}={\bf 0},+}^{\tilde\Lambda^2=3}  & = &
\left(
\begin{array}{ccccc}
-2\mu & 2x & 0 & 0  \\
2x & 1 &  \sqrt{2} x  & 0  \\
0 &  \sqrt{2} x  & 2+2\mu &  \sqrt{2} x \\
0 & 0 &  \sqrt{2} x  & 3   \\
 \end{array}
 \right)
 \ =\
 {3\over 2} I_4\ +\
 \left(
\begin{array}{ccccc}
-{3\over 2}-2\mu & 2x & 0 & 0  \\
2x & -{1\over 2} &  \sqrt{2} x  & 0  \\
0 &  \sqrt{2} x  & {1\over 2}+2\mu &  \sqrt{2} x \\
0 & 0 &  \sqrt{2} x  & {3\over 2}   \\
 \end{array}
 \right)
 \ =\
  {3\over 2} I_4\ +\ H_\Tslash
 \ \ ,
\end{eqnarray}
which has been split into a term proportional to the identity matrix and a traceless term.
The term proportional to the identity matrix is dropped until the end of the calculation, as it contributes only an overall phase,  and we focus on the traceless matrix
$H_\Tslash$.
With example values of Hamiltonian parameters, $\mu=0.1$ and $x=0.6$, this matrix has eigenvalues
$E_{\Tslash i} = -2.51164, \ -0.397399,\ 0.768049, \ 2.14099$.
$H_\Tslash$ can be projected onto the generators of $SU(4)$ transformations in the preferred basis
\begin{eqnarray}
H_\Tslash & = & \sum_i\ \ c_i\ {\cal O}_i
\ \ ,
\end{eqnarray}
where the operator basis is defined to be
\begin{eqnarray}
{\cal O}_1 & = & \sigma_x\otimes\sigma_x
\ =\
\left(
\begin{array}{cccc}
0&0&0&1 \\ 0&0&1&0 \\ 0&1&0&0 \\ 1&0&0&0
\end{array}
\right)
\ ,\
{\cal O}_2 \ =\  \sigma_x\otimes\sigma_y
\ =\
\left(
\begin{array}{cccc}
0&0&0&-i \\ 0&0&i&0 \\ 0&-i&0&0 \\ i&0&0&0
\end{array}
\right)
\nonumber\\
{\cal O}_3 & = & \sigma_x\otimes\sigma_z
\ =\
\left(
\begin{array}{cccc}
0&0&1&0 \\ 0&0&0&-1 \\ 1&0&0&0 \\ 0&-1&0&0
\end{array}
\right)
\ ,\
{\cal O}_4 \ =\ \sigma_y\otimes\sigma_x
\ =\
\left(
\begin{array}{cccc}
0&0&0&-i \\ 0&0&-i&0 \\ 0&i&0&0 \\ i&0&0&0
\end{array}
\right)
\nonumber\\
{\cal O}_5 & = & \sigma_y\otimes\sigma_y
\ =\
\left(
\begin{array}{cccc}
0&0&0&-1 \\ 0&0&1&0 \\ 0&1&0&0 \\ -1&0&0&0
\end{array}
\right)
\ , \
{\cal O}_6 \ =\ \sigma_y\otimes\sigma_z
\ =\
\left(
\begin{array}{cccc}
0&0&-i&0 \\ 0&0&0&i \\ i&0&0&0 \\ 0&-i&0&0
\end{array}
\right)
\nonumber\\
{\cal O}_7 & = & \sigma_z\otimes\sigma_x
\ =\
\left(
\begin{array}{cccc}
0&1&0&0 \\ 1&0&0&0 \\ 0&0&0&-1 \\ 0&0&-1&0
\end{array}
\right)
\ ,\
{\cal O}_8 \ =\  \sigma_z\otimes\sigma_y
\ =\
\left(
\begin{array}{cccc}
0&-i&0&0 \\ i&0&0&0 \\ 0&0&0&i \\ 0&0&-i&0
\end{array}
\right)
\nonumber\\
{\cal O}_9 & = & \sigma_z\otimes\sigma_z
\ =\
\left(
\begin{array}{cccc}
1&0&0&0 \\ 0&-1&0&0 \\ 0&0&-1&0 \\ 0&0&0&1
\end{array}
\right)
\ \ ,
\end{eqnarray}
and
\begin{eqnarray}
{\cal O}_{10} & =& I\otimes\sigma_x
\ =\
\left(
\begin{array}{cccc}
0&1&0&0 \\ 1&0&0&0 \\ 0&0&0&1 \\ 0&0&1&0
\end{array}
\right)
\ , \
{\cal O}_{11}\ =\  I\otimes\sigma_y
\ =\
\left(
\begin{array}{cccc}
0&-i&0&0 \\ i&0&0&0 \\ 0&0&0&-i \\ 0&0&i&0
\end{array}
\right)
\nonumber\\
{\cal O}_{12} & = &   I\otimes\sigma_z
\ =\
\left(
\begin{array}{cccc}
1&0&0&0 \\ 0&-1&0&0 \\ 0&0&1&0 \\ 0&0&0&-1
\end{array}
\right)
\ ,\
{\cal O}_{13} \ =\ \sigma_x \otimes I
\ =\
\left(
\begin{array}{cccc}
0&0&1&0 \\ 0&0&0&1 \\ 1&0&0&0 \\ 0&1&0&0
\end{array}
\right)
\nonumber\\
{\cal O}_{14} & = &  \sigma_y \otimes I
\ =\
\left(
\begin{array}{cccc}
0&0&-i&0 \\ 0&0&0&-i \\ i&0&0&0 \\ 0&i&0&0
\end{array}
\right)
\ ,\
{\cal O}_{15} \ =\ \sigma_z \otimes I
\ =\
\left(
\begin{array}{cccc}
1&0&0&0 \\ 0&1&0&0 \\ 0&0&-1&0 \\ 0&0&0&-1
\end{array}
\right)
\ \ .
\end{eqnarray}
The operators are normalized such that
${\rm Tr}\left[\ {\cal O}^\dagger_{i} {\cal O}_{j}  \right] = 4 \delta_{ij}$.
Performing traces gives
\begin{eqnarray}
c_1 & = & c_5 \ =\ {x\over\sqrt{2}}\ =\ 0.424264
\ \ ,\ \
c_7\ =\ x\left(1-{1\over\sqrt{2}}\right)\ =\ 0.1757359
\nonumber\\
c_9 & = &  -\mu\ =\ -0.1
\ \ ,\ \
c_{10}\ =\ x\left(1+{1\over\sqrt{2}}\right)\ = \ 1.024264
\nonumber\\
c_{12} & = & -{1\over 2}
\ \ ,\ \
c_{15}\ =\ -\left(1+\mu\right)\ =\ -1.100
\ \ .
\end{eqnarray}

As
phase re-definitions of the four eigenstates can be performed, the symmetry group relevant to the variational calculations involving the $4\times 4$ Hamiltonian is $SO(4)$ (with six generators).
Starting from the orthonormal basis of states
$\{  (1,0,0,0)^T , (0,1,0,0)^T , (0,0,1,0)^T , (0,0,0,1)^T \}$,
 values of the six angles  that   diagonalize the Hamiltonian matrix, $H_\Tslash$ are required.
Given the nearest-neighbour structure of  $H_\Tslash$,
the ground states is of the form
\begin{eqnarray}
S(\theta_1,\theta_2,\theta_3)_{gs}
& = &
R_{34}(\theta_3)\
R_{23}(\theta_2)\
R_{12}(\theta_1)
\ (1,0,0,0)^T
\ \ ,
\label{eq:variationalwavefunction}
\end{eqnarray}
where
\begin{eqnarray}
R_{12}(\theta) & = &
\left(
\begin{array}{cccc}
\cos\theta &-\sin\theta&0&0 \\ \sin\theta&\cos\theta&0&0 \\ 0&0&1&0 \\ 0&0&0&1
\end{array}
\right)
\ \ ,\ \
R_{23}(\theta) \ =\
\left(
\begin{array}{cccc}
1& 0&0&0\\
0&\cos\theta &-\sin\theta&0 \\ 0&\sin\theta&\cos\theta&0 \\ 0&0&0&1
\end{array}
\right)
\nonumber\\
R_{34}(\theta) & = &
\left(
\begin{array}{cccc}
1& 0&0&0\\
0&1& 0&0\\
0&0&\cos\theta &-\sin\theta \\ 0&0&\sin\theta&\cos\theta
\end{array}
\right)
\ \ .
\end{eqnarray}
An exact minimization (Mathematica) gives
$\theta_1=-0.6130$, $\theta_2 = -0.2785$ and $\theta_3=-0.20844$.
Applying this transformation to the other vectors produces three orthonormal vectors that are orthogonal to the ground state and form a basis for the excited states.  The resulting Hamiltonian in that sector is also traceless and contains only nearest-neighbor interactions, making the variational determination of excited states significantly less costly than methods recently proposed for the determination of eigenstates without this simple structure \cite{Santagati2018}. A similar form of the variational wavefunction to Eq.~\eqref{eq:variationalwavefunction} involving only two angles
can be used to construct the first excited state.  The same procedure can be repeated to obtain all eigenstates.

The expectation value of the energy in any given state defined by the angles $\theta_i$ is
\begin{eqnarray}
\langle H_\Tslash \rangle_{\theta_i} & = &
(1,0,0,0)R_{12}(\theta_1)^T R_{23}(\theta_2)^{T} R_{34}(\theta_3)^{T}
H_\Tslash R_{34}(\theta_3) R_{23}(\theta_2) R_{12}(\theta_1) (1,0,0,0)^T
\nonumber\\
& = & \sum_i \
c_i \ (1,0,0,0)R_{12}(\theta_1)^T R_{23}(\theta_2)^{T} R_{34}(\theta_3)^{T}
{\cal O}_i
R_{34}(\theta_3) R_{23}(\theta_2) R_{12}(\theta_1) (1,0,0,0)^T
\ \ ,
\end{eqnarray}
and therefore the expectation values $\langle {\cal O}_i \rangle_{\theta_i} $ need to be calculated to form $\langle H_\Tslash \rangle_{\theta_i} $, which is then extremized to determine the angles in the wavefunction.
The operators ${\cal O}_{9,12,15}$
are diagonal from the circuit used to determine the time-dependence of the pair-production,
while the other operators require additional gates to transform into a diagonal basis in preparation for measurement:
\begin{eqnarray}
 \mathcal{O}_1 & : & {\rm H}\otimes {\rm I} \  {\rm I} \otimes {\rm H}  |q_0 q_1 \rangle
 \ \ ,\ \
 \mathcal{O}_1(diag) \ =\ {\rm diag}(1,-1,-1,1)
 \nonumber\\
  \mathcal{O}_5 & : & {\rm H}\otimes  {\rm I} \    {\rm S}^\dagger\otimes  {\rm I}  \   {\rm I} \otimes {\rm H} \   {\rm I} \otimes {\rm S}^\dagger |q_0 q_1 \rangle
 \ \ ,\ \
 \mathcal{O}_5(diag) \ =\ {\rm diag}(1,-1,-1,1)
 \nonumber\\
\ \ ,
  \mathcal{O}_{7} & : &   {\rm I} \otimes {\rm H}  |q_0 q_1 \rangle
 \ \ ,\ \
 \mathcal{O}_{7}(diag) \ =\ {\rm diag}(1,-1,-1,1)
 \nonumber\\
  \mathcal{O}_{10} & : &   {\rm I} \otimes {\rm H}  |q_0 q_1 \rangle
 \ \ ,\ \
 \mathcal{O}_{10}(diag) \ =\ {\rm diag}(1,-1,1,-1)
\end{eqnarray}
An initial grid-based sampling of approximately 10 sets of angles for the low-depth circuit of Eq.~\eqref{eq:variationalcircuit} is used with a set of uniform Bayesian priors to establish a posterior distribution for the three angles.  A second iteration of the process yields a sufficiently precise determination of the ground state energy.
\begin{equation}
  \begin{gathered}
\Qcircuit @C=-0.7em @R=-0.5cm @! {
  & \gate{e^{-i \theta_1 \frac{\sigma_y}{2}}} &  \ctrl{1} & \qw & \ctrl{1} & \qw & \qw\\
  &  \gate{e^{-i \theta_0 \frac{\sigma_y}{2}}} & \targ & \gate{e^{-\theta_0\frac{\sigma_y}{2}}} & \targ & \gate{e^{-i\theta_2 \frac{ \sigma_y}{2}}}& \qw
}
\end{gathered}
\label{eq:variationalcircuit}
\end{equation}

%%%%%%%%%%%%%%%%%%%%%%%%%%%%%%%%%%%%%%%%%%%%%%%%%%
\section{Convergence with the cut-off in the gauge-field energy}
\label{app:convergence}
\begin{figure}[!th]
\includegraphics[width=0.8\textwidth]{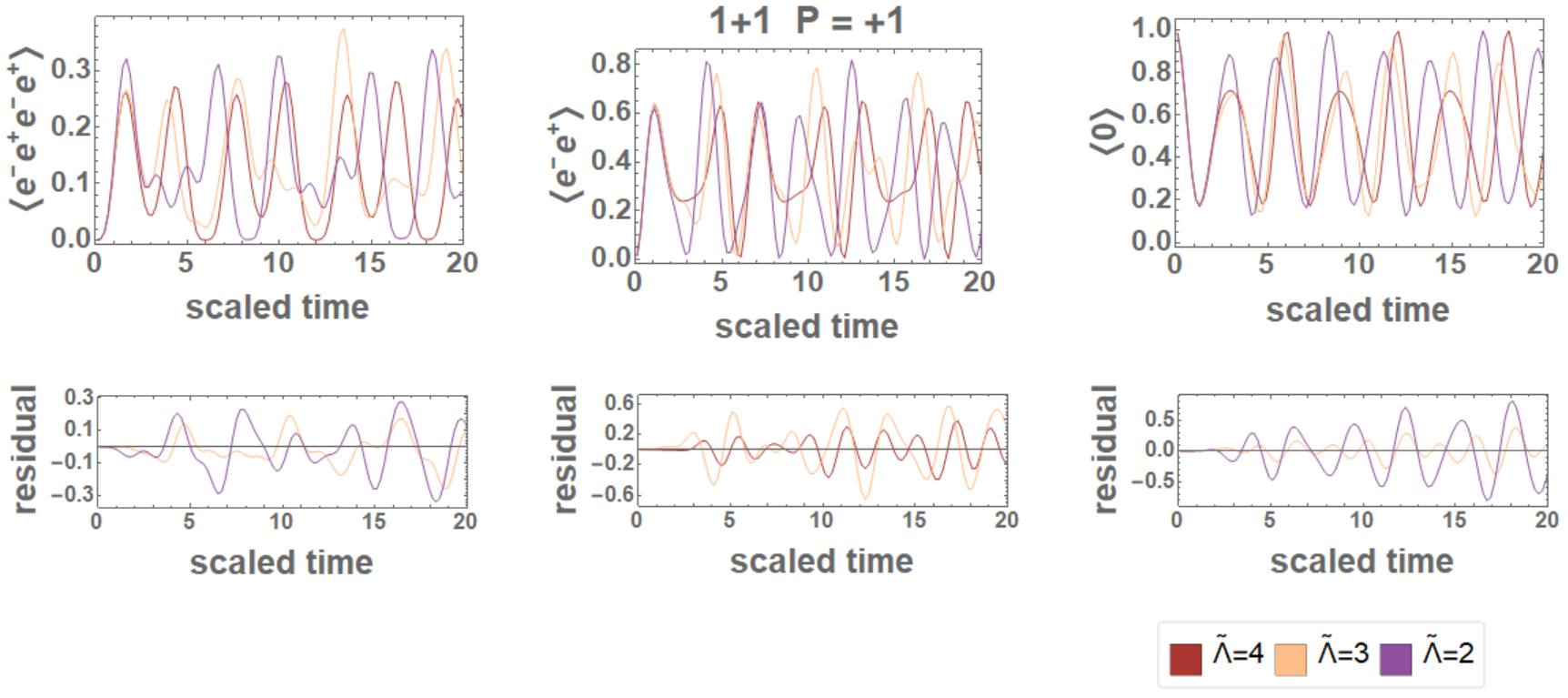}
\includegraphics[width=0.8\textwidth]{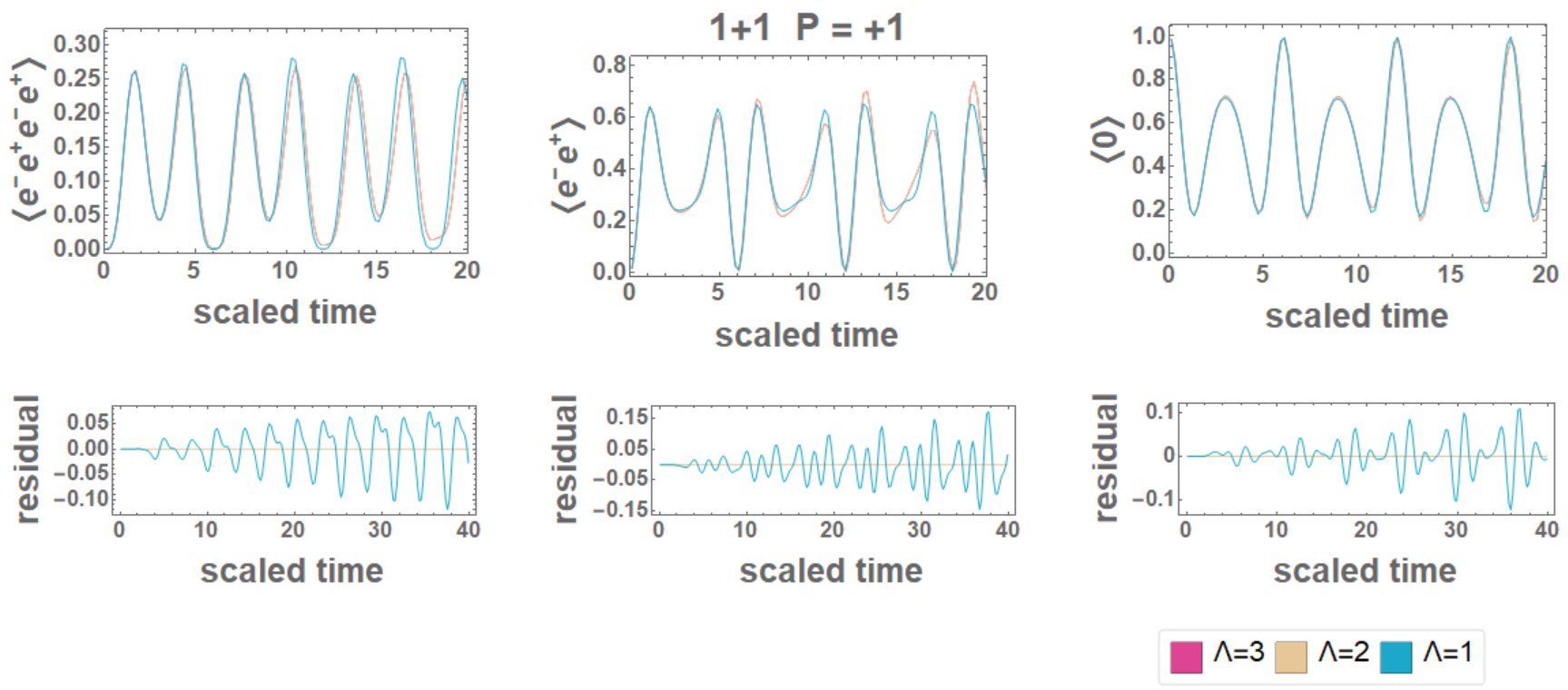}
\caption{
The upper six panels show the convergence of the dynamical pair fluctuations with increasing energy truncation in the electric field,
$\tilde\Lambda$,
with the cut off in the energy in each electric flux link $\Lambda = 1$.
With so few states present, significant modifications are seen with each value of $\tilde{\Lambda}$.
The second row of panels show the residuals
of the upper row from the untruncated value of $\tilde{\Lambda} = 4$.
The lower six panels show early convergence in $\Lambda$.  The dynamics are found to be stable to the introduction of high-energy states beyond $\Lambda=2$.}
\label{fig:dyamicsconvergence}
\end{figure}

\FloatBarrier
\begin{table}
\centering
\begin{tabular}{lccccc}
\hline
Even parity &GS & E1 & E2 & E3 & E4 \\
\hline
\hline
Exact & -1.0118 & 1.0771 & 2.0966 & 3.1037 & 4.3044 \\
\hline
\hline
$\tilde{\Lambda}=4$ & -1.0118 & 1.0784 & 2.1120 & 3.1666 & 4.4549 \\
$\tilde{\Lambda}=3$ & -1.0116 & 1.1026 & 2.2681 & 3.6410 & - \\
$\tilde{\Lambda}=2$ & -1.0076 & 1.2440 & 2.7635 & - & - \\
$\tilde{\Lambda}=1$ & -0.9416 & 1.7416 & - & - & - \\
 \hline
 \end{tabular}\quad\quad\quad\quad\quad\quad
\begin{tabular}{lcccc}
\hline
Odd parity&GS & E1 & E2 & E3\\
\hline
\hline
Exact & 0.4857 & 1.9149 & 3.0670 & 4.3025 \\
\hline
\hline
$\tilde{\Lambda}=4$ & 0.4859 & 1.9281 & 3.1323 & 4.4536 \\
$\tilde{\Lambda}=3$ & 0.4929 & 2.0816 & 3.6254 & - \\
$\tilde{\Lambda}=2$ & 0.5608 & 2.6392 & - & - \\
\hline
\vspace{0.03cm}
\end{tabular}
\caption{
The classically determined energy spectra of the low-energy $\mathbf{k} = \mathbf{0}$ Hilbert space further truncated
by the total energy in the electric field,
$\tilde{\Lambda}$, the largest value of $\sum_i \ell_i^2$ retained in the space.
Reducing this cutoff sequentially removes the highest energy state from the basis, as shown by the rows in each table. In the $P=+1$ sector with $\tilde{\Lambda} = 3$ (the reduced 2-qubit form),
the systematic error in the ground state energy introduced by this truncation
is less than $1\%$.
}
\label{tab:energyconvergence}
\end{table}
While the implementation of the constraints imposed by Gauss's law, momentum projections, and parity projections reduce the size of the Hilbert space of the 1+1 system sufficiently to permit calculations on  3-qubit and 2-qubit quantum computers,
a further truncation of the total energy in the electric field $\tilde{\Lambda} = \sum\limits_i \ell_i^2$ allows approximate calculations on even smaller numbers of qubits.  Table~\ref{tab:energyconvergence} shows the
classically calculated (Mathematica)
convergence of the energy spectrum as a function of the cutoff $\tilde{\Lambda}$.
By removing the highest-energy state, the system retains the value of the ground state at a precision of better than 1\%. It is then pertinent to also ask about the convergence of dynamical properties with the cutoff $\tilde{\Lambda}$.
Figure~\ref{fig:dyamicsconvergence}
shows the results of classical (Mathematica) calculations of the probabilties of finding $e^+e^-$ pairs at some time after initializing the system. Rapid convergence is found in raising the energy cut-off associated with each electric flux link, and convergence is also found in raising the total allowed energy in the electric field only once $\Lambda$ has been chosen large enough.
The upper row of plots has been constructed with a per-link cutoff of $\Lambda=1$ and further reductions in the total energy, $\tilde{\Lambda}$, resulting in significant modifications with each value of $\tilde{\Lambda}$. As this value of $\Lambda$ was chosen for its $\sim 1\%$ errors on the ground state energy, it may have been tempting to think that also the dynamics are converged at this level of truncation.  However, because a well-reproduced ground state energy is a relatively weak constraint on the exact form of the wavefunction,  a truncation leading to a precise calculation of the energy may be insufficient to accurately capture dynamics.
It can be seen from the next row of Fig.~\ref{fig:dyamicsconvergence} that, even without a $\tilde{\Lambda}$ cutoff, this system with $\Lambda=1$ does not yet have converged dynamics.
This convergence of ground state properties before dynamics has practical implications for the preparation of quantum states using ground state explorations such as VQE.

In addition to giving confidence in the accuracy and precision of the calculations performed on the quantum hardware, it also suggests a means to improve the variational methods applied to these particular calculations.
The number of angles required to specify the ground state is smaller for a lower energy cut off.  As such, the Bayesian priors associated with the angles in the variational ansatz provide a perturbatively close set of priors for a subset of angles in systems with larger energy  cut offs.  This hierarchy has been explicitly verified.

%%%%%%%%%%%%%%%%%%%%%%%%%%%%%%%%%
\section{Scaling to Larger Lattices}
\label{app:largerlattices}
\FloatBarrier

By determining the physical subspace and projecting onto states of zero momentum and definite parity, the dimensionality of the Hilbert space is exponentially reduced.  In its original latticized form with 1 qubit for every site and two qubits for every link ($\Lambda = 1$), the Hilbert space grows with $N_\text{s}$, the number of spatial sites, as $e^{\log(64)N_\text{s}}$.  By enforcing the local constraint of Gauss's law, this exponent is significantly reduced to   $1.02(1)e^{1.1772(2)N_\text{s}}$.
With further projection to $\mathbf{k}=\mathbf{0}$ and even-parity, the scaling of the relevant Hilbert space becomes $0.29(5)e^{1.006(23) N_\text{s}}$.
The coefficients and exponents have been determined by fitting the numerically-calculated dimensions
given in Table~\ref{tab:scaling} on the scaling of $D_\text{physical}$.   This is achievable through combinatoric calculations of a non-trivial binary tree at and beyond 80 spatial sites.
Similar combinatoric methods remain to be devised for $D_{{\bf k}={\bf 0}}$ and $D_{\text{even/odd}}$ due to the additional complexity of global symmetry constraints identified between entire branches of the tree structure.
With each reduction, an exponentially large unphysical or symmetry-disconnected contribution to the Hilbert space is removed, eliminating the possibility of introducing errors associated with propagating states into these undesirable regions.

\begin{table}[!th]
\centering
\begin{tabular}{ccc|c|c|cccc}
\hline
\ \  \text{physical sites}\ \  & \ \ ${Nq}_{\text{lattice}}$\ \  &\ \  $D_{\text{lattice}}$\ \  & \ \ $D_{\text{physical}}$\ \ & \ \ $D_{{\bf k}={\bf 0}}$ \ \  & \ \ $D_{\text{even}}$\ \  &\ \  $D_{\text{odd}}$ \ \  & $\text{Nq}_{\text{even}}^{\mathbf{k}=\mathbf{0}}$ & $\text{Nq}_{\text{odd}}^{\mathbf{k}=\mathbf{0}}$ \\
 \hline
 \hline
 1 & 6 & 64 & 5 & - & 3 & 2 & 2 & 1 \\
 2 & 12 & $4.1\times10^3$ & 13 & 9 & 5 & 4 & 3 & 2 \\
 4 & 24 & $1.7\times10^7$ & 117 & 35 & 19 & 16 & 5 & 4 \\
 6 & 36 & $6.9\times10^{10}$ & 1,186 & 210 & 110 & 100 & 7 & 7 \\
 8 & 48 & $2.8\times10^{14}$ & \text{12,389} & 1,569 & 801 & 768 & 10 & 10 \\
 10 & 60 & $1.2\times10^{18}$ & \text{130,338} & 13,078 & 6,593 & 6,485 & 13 & 13 \\
 12 & 72 & $4.7\times10^{21}$ &\  \text{1,373,466}\  &\  114,584\  & 57,468 & 57,116 & 16 & 16 \\
 \hline
\end{tabular}
\caption{Scaling of the Hilbert space with different levels of reduction and projection.
Moving from left to right, the qubit mapping begins with the lattice through which the Schwinger model is naturally defined,
is constrained by Gauss' law to allow only physical states,
is projected to zero momentum configurations,
and finally projected onto states of definite parity.
The number of required qubits grows linearly in the size of the system both before and after the reduction,
however this reduction decreases the coefficient of this linear scaling from 6 to 1.27(5).}
\label{tab:scaling}
\end{table}

%%%%%%%%%%%%%%%%%%%%%%%%%%%%%%%%%
\section{Quantifying the CNOT systematic errors}
\label{app:CNOTerrors}

\begin{figure}[!ht]
  \includegraphics[width = 0.65\textwidth]{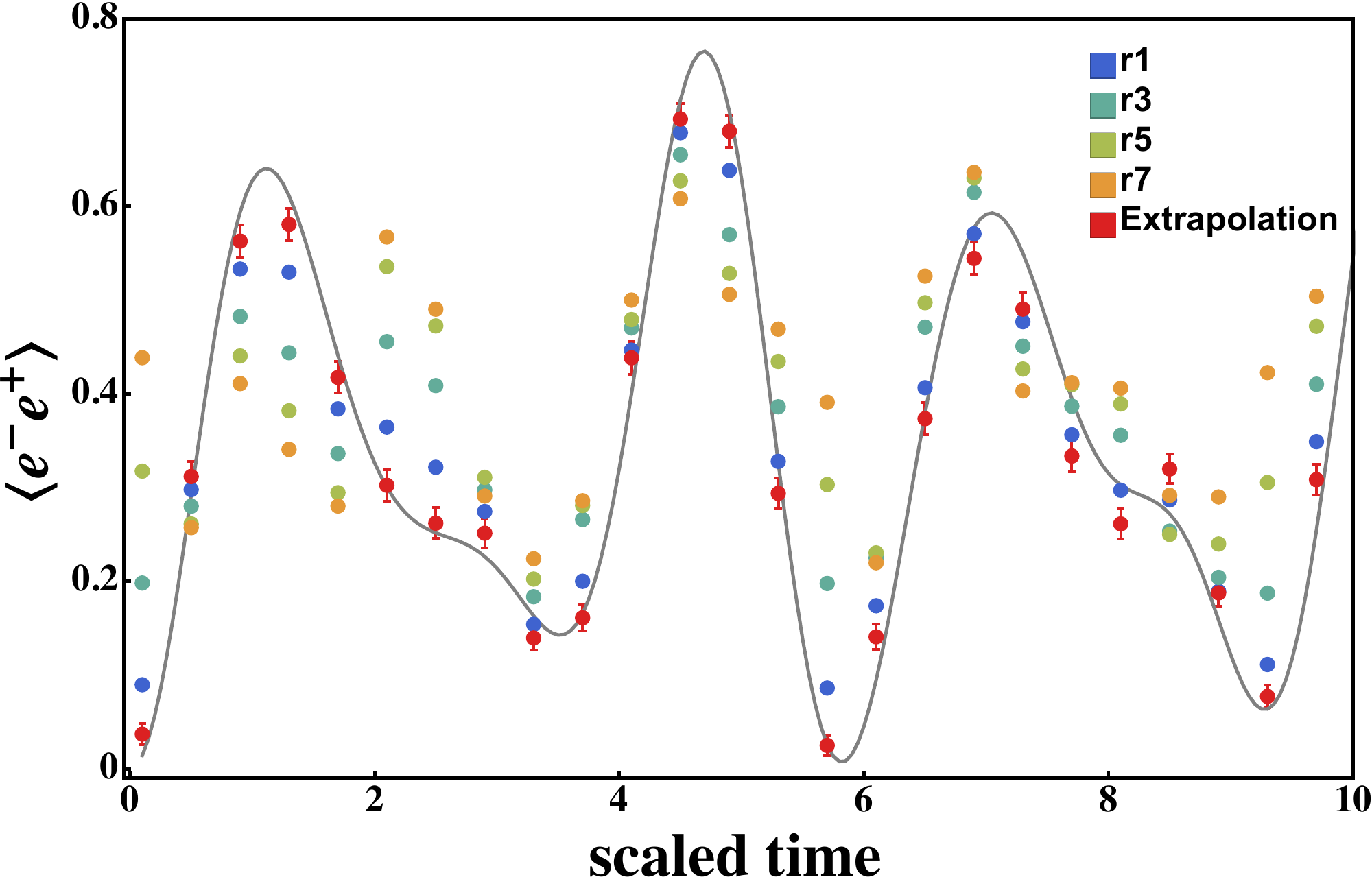}
  \caption{
  The single $e^+e^-$ pair density in the ground state of the $1+1$, two-spatial site Schwinger model as a function of time starting from the empty vacuum, calculated with different numbers of CNOT gates.
  A quadratic extrapolation in the CNOT-gate systematic error has been performed---shown by the red points (those with visible error bars).
  The results shown here were determined with 8K measurements per point.
  The exact result is given by the solid gray curve.
  (500 IBM allocation units were used for the~$\sim6.1$~QPU$\cdot$s needed to generate this data set.)
  }
  \label{fig:CNOTextrapPairs}
\end{figure}

The most significant systematic uncertainties we encountered in executing quantum circuits on the IBM quantum computing hardware (ibmqx2 and ibmqx5) were introduced by CNOT gates, as is well known, see for example, Ref.~\cite{kandala2017,PhysRevLett.120.210501}.
In order to quantify and remove this systematics from the dynamics of calculated observables, a series of additional calculations were performed in which each single CNOT gate in a circuit was replaced by an odd-number of CNOT gates, ranging from $r=1,3,5,7$ gates at each insertion (and  up to 25 CNOT gates in some exploratory cases).  These measurement results were then used to perform an extrapolation to $r=0$. To model the process, we assumed that each ideal CNOT operation is followed by a depolarizing two-qubit channel (white noise model) resulting in a fractional CNOT error $\epsilon_g$ associated with each CNOT gate. Applying the CNOT gate $r$ times results in the output density matrix $\rho_{out} = (1-r\epsilon_g) {\rm CNOT} \rho_{in} {\rm CNOT} + r\epsilon_g I + \mathcal{O}(\epsilon_g^2)$ where we used the fact that CNOT$^2=I$ and that it commutes with the white noise channel. Therefore, the expectation value of any Pauli operator $O$ measured after $r$ noisy CNOT application will relate to its "noiseless" ($r=0$) value through a linear equation $\expect{O}(r) = \expect{O}(0) - r\expect{O}(0)\epsilon_g$ for small values of $\epsilon_g$~\cite{Li2017}. Next, linear and quadratic fits in $\epsilon_g$ were performed on each temporal ensemble of data. The results obtained through such extrapolations for the time-dependence of the $e^+e^-$ pair density in the vacuum of the $1+1$, two-spatial-site Schwinger model are shown in Fig.~\ref{fig:CNOTextrapPairs}.  In this figure, quadratic extrapolation in $r$ is seen to be crucial in calculating the true dynamic evolution of pair production.

\begin{figure}[!ht]
  \includegraphics[width = 0.47\textwidth]{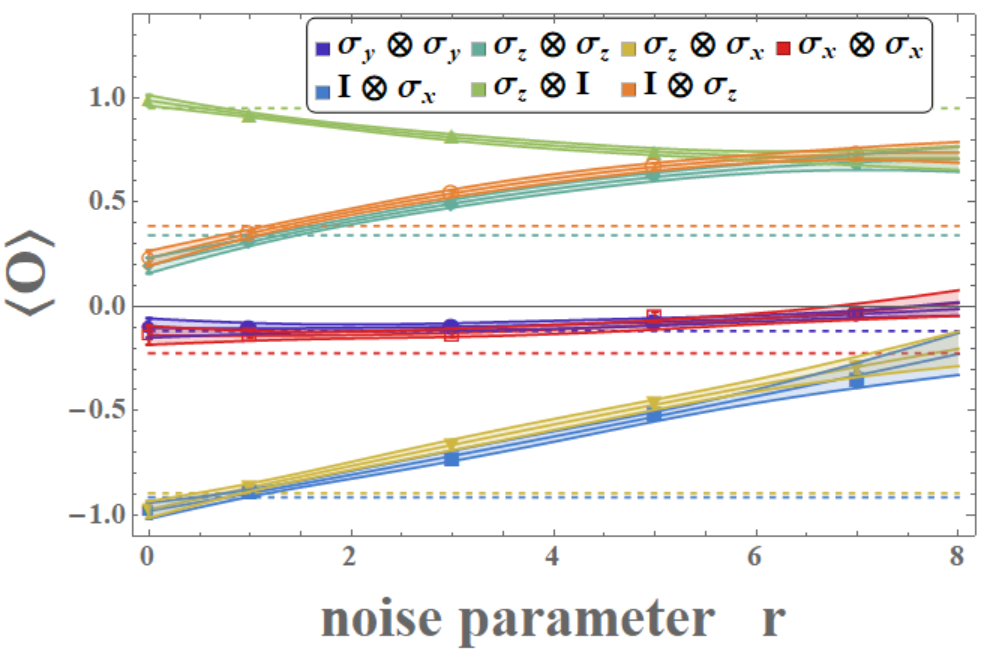}
  \includegraphics[width = 0.47\textwidth]{Hp3_HandCC.png}
  \caption{
  The left panel shows the expectation values of operators contributing to the Hamiltonian and the chiral condensate,
  as described in the text,
  at angles (see Eq.~\eqref{eq:variationalcircuit}) describing the variational energy minimum of the $r=1$ system.
  The number of CNOT gates (the ``noise parameter'' $r$) is swept through $r=1,3,5,7$ wherever one appears in the circuit (Eq.~\eqref{eq:variationalcircuit}).
  The right panel shows the ground state energy and chiral condensate at the variational ground state (purple, blue extrapolated to -1.000(65) and -0.296(13), respectively).  Points at $r = 0$ have been quadratically extrapolated to remove this systematic bias while the horizontal dashed lines indicate the exact values. (1200 IBM allocation units were used for the $\sim6.35$~QPU$\cdot$s needed to generate this data set.)
  }
  \label{fig:CNOTvar}
\end{figure}

Applying this method of CNOT error extrapolation to the variational calculations of the operator expectation values, ground-state energy and chiral condensate demands more care.  This can be seen in Fig.~\ref{fig:CNOTvar} where a Bayesian optimization has been performed to find the 3 angles in Eq.~\eqref{eq:variationalcircuit} that minimize the calculated energy using the original circuit ($r=1$).  These three angles are then used to implement 10 samples of the operator expectation values (ibmqx5, 8192 shots) at increased values of the bias (increased $r$). The results of this procedure are then fit to a quadratic form in $r$ with confidence intervals representing 68\%  on the mean value under the assumption of only Gaussian fluctuations.

The reason additional care is needed when applying this CNOT extrapolation to ground state searches as opposed to the dynamic evolution of Fig.~\ref{fig:CNOTextrapPairs} is due to the inherent bias when optimizing with the original circuit ($r = 1$).  Removing this bias and optimizing the angles used to implement the circuit evolving to the ground state are not commuting actions.  This can be intuitively understood by regarding the introduction of additional CNOTs (and their associated systematics) as non-unitary contributions to the evolution.  As such, the energy hypersurface that the angles minimize has itself been modified.  Performing the extrapolation in $r$ as has been done in Fig.~\ref{fig:CNOTvar} steps us into the correct energy landscape ($r = 0$) but does not send the calculation to the $r = 0$ ground state.  This can be seen numerically in the deviation of many operator terms from $r = 1$ to $r = 0$ away from the true values.  However, calculated values of the energy and the chiral condensate on the $r=0$ hypersurface with angles optimized at $r=1$ are consistent with expected values, supporting the expectation of low-order-polynomial extrapolation between hypersurfaces. This further indicates that the assumed white noise model is valid only approximately and better experimental characterization of the noise processes is needed.

In order to extrapolate to the $r = 0$ ground state, the minimization and extrapolation procedures must be interchanged so that the Bayesian optimization is performed on the $r = 0$ hypersurface of interest.  Inverting the extrapolation and optimization in this way would increase the cost of the variational method by roughly a factor of 4 (the number of $r$ values needed for a meaningful extrapolation to $r = 0$) but will allow the ground state wavefunction to be determined with the CNOT bias removed.  In this way, extrapolations of the systematic error associated with CNOT gates will be essential in obtaining physical results of scientific accuracy.

The linear and quadratic analyses we have performed are appropriate when the systematic errors from the CNOT gates remain small.
However, for a sufficiently large value of $r$, and beyond, significant non-linearities will become important, and in particular the transition to the classical regime will render values of observables independent of $r$.
\begin{figure}[!ht]
  \includegraphics[width = 0.65\textwidth]{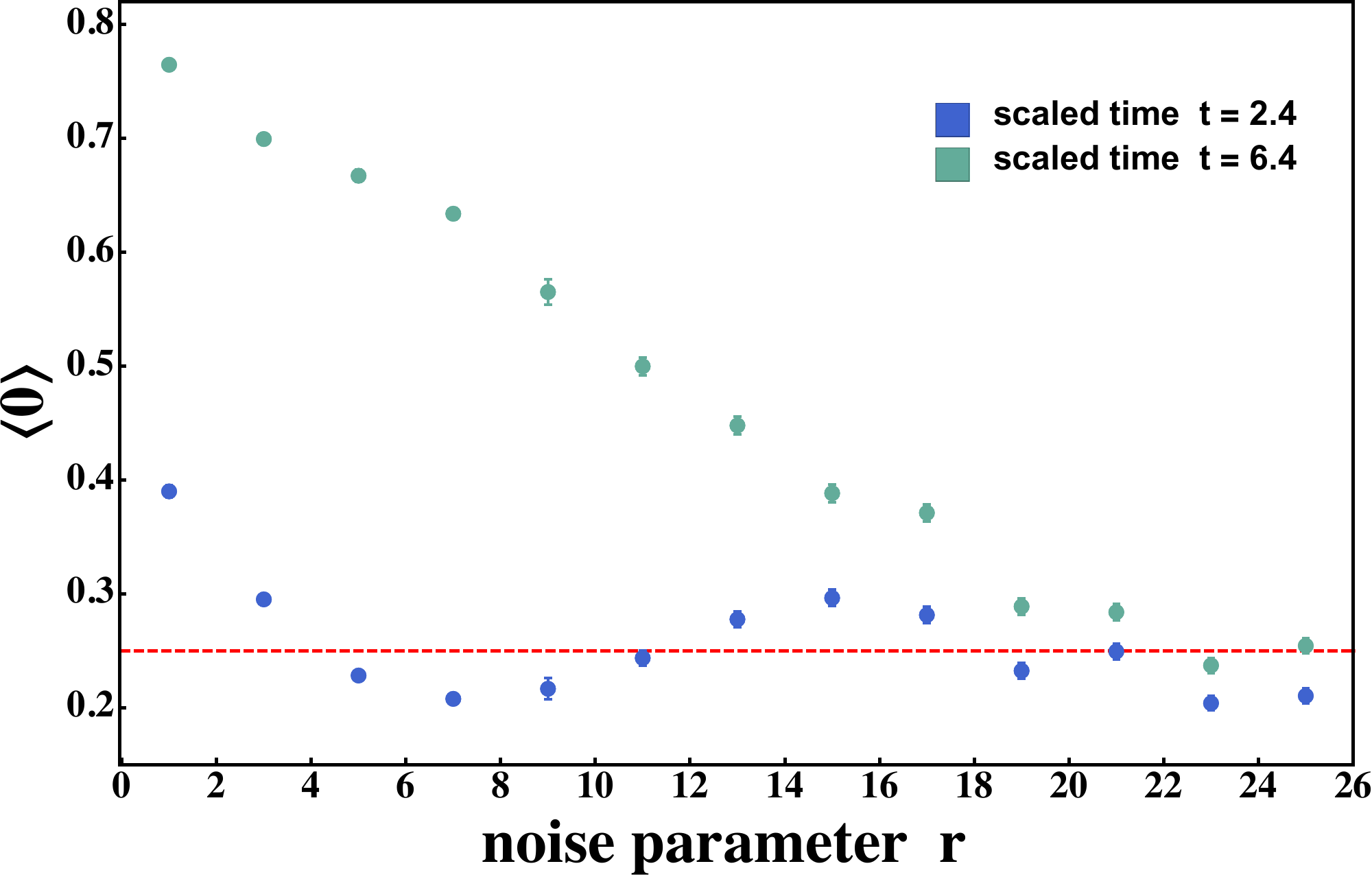}
  \caption{
  The behavior of the CNOT-gate systematic errors
  in the probability of finding zero $e^-e^+$ pairs
  as a function of the number of CNOT gates for times $t=2.4$ and $t=6.4$ (lower and upper points, respectively) in the evolution of the $1+1$, two spatial site Schwinger model.
  The red dashed line corresponds to the classical value of $0.25$.
  (130 IBM allocation units were used for the $\sim4.6$~QPU$\cdot$s needed to generate this data set.)
  }
  \label{fig:rCNOTextrap0Pairs}
\end{figure}
Figure~\ref{fig:rCNOTextrap0Pairs} shows the probability of finding zero pairs in the ground state at $t=2.4$ and $t=6.4$ as a function of the CNOT-gate depth per insertion point.  A linear form for an extrapolation to zero error is valid only for small gate depth at $t=2.4$, but for a much larger gate depth at $t=6.4$.
For large CNOT gate counts, an oscillatory behavior in $r$ is observed at $t=2.4$, while at other times, the situation is less severe.  Note that the time scale for the system to approach this classical limit (where the density matrix tends to the identity) is much greater than that explored in Fig.~\ref{fig:CNOTvar} and explains the observation that three of the seven operators have not yet been driven to zero by $r = 7$.

%%%%%%%%%%%%%%%%%%%%%%%%%%%%%%%%%
\section{The chiral condensate  $\langle\overline{\psi}\psi\rangle$}
\label{app:chiralcondensate}

In nature,
the QCD chiral condensate of the vacuum plays a critical role in determining the nature of low-energy strong interactions.  Its non-zero value spontaneously breaks the approximate chiral symmetries of the QCD Lagrange density, leading to three light pseudo-Goldstone bosons, the pions, which are responsible for the long-range component of the nuclear force.
In the $0+1$ and $1+1$ Schwinger model, the ground state also has a non-zero value for the chiral condensate $\langle\overline{\psi}\psi\rangle$ for the values of parameters we have chosen to analyze.
The chiral condensate provides a different probe of the structure of the ground states, beyond what is revealed by its absolute energy density.

At the level of fermion sites, the chiral condensate operator is given by
\begin{eqnarray}
\hat\chi_0 & = &
{1\over N}
\left(
\sum_{i=odd}^{N_Q}\hat S_z^{(i)} - \sum_{i=even}^{N_Q}\hat S_z^{(i)}
\right)
 \ \ ,
\end{eqnarray}
where $N$ is the number of fermion sites in the system.
In the antiferromagnetic state (the strong-coupling ground state), $\langle \hat\chi_0 \rangle = -{1\over 2}$.
In the two-qubit bases we have been working with to describe the dynamics of the $1+1$, two-spatial-site even-parity sector, this operator has a matrix representation,
\begin{eqnarray}
\hat\chi_0 & \rightarrow &
{1\over 2}\
\left(
\begin{array}{cccc}
-1&0&0&0\\
0&0&0&0 \\
0&0&1&0 \\
0&0&0&0
\end{array}
\right)
\ =\
-{1\over 4}
\left(\
\sigma_z\otimes\sigma_z\ +\ \sigma_z\otimes I_2
\ \right)
 \ \ .
\end{eqnarray}
Including this operator in the variational calculation of the ground state energy, which can be done easily as the operators contribute to both quantities, produces a value of $\langle\overline{\psi}\psi\rangle=-0.296(13)$ that is consistent with the exact known result, as shown in Table~\ref{tab:vacuum}.
It is interesting to observe that the value of the condensate varies more strongly near the ground state energy minimum than the energy does. This is not a surprise given that it is sensitive to different attributes of the ground state than the energy.

%%%%%%%%%%%%%%%%%%%%%%%%%%%%%%%%%
\section{Code Snippets used for Calculations on IBM Hardware}

It may be of benefit to the reader who wishes to reproduce our results to have snippets of the
Python3 scripting that created the circuits that were executed on the IBM simulators and hardware.
The following lines of code were written by one of the co-authors (Savage), and produced results that were verified by
analogous (independent) scripts written by multiple other co-authors.
To determine the $e^+e^-$ pair density in the ${\bf k}={\bf 0}$ even-parity sector of the two spatial site Schwinger model, a list of sets of nine angles were determined classically (Mathematica) which were input to the Python3 script (into a list called {\tt angletab}).
The circuits were created with the following {\tt for} loop:
\begin{verbatim}
for ii in range(0,len(angletab)):
        p0=qp.get_circuit(pidtab[ii])
        angles = angletab[ii]
        print("Calculating angles ii = ",ii," : = ",angles)

        a1=angles[0]
        a2=angles[1]
        a3=angles[2]
        a4=angles[3]
        a5=angles[4]
        a6=angles[5]
        a7=angles[6]
        a8=angles[7]
        a9=angles[8]

# acting with Kp

        p0.u3(a2,0,0,qr[1])
        p0.u3(0,0,a3,qr[1])
        p0.u3(a5,0,0,qr[0])
        p0.u3(0,0,a6,qr[0])

# acting with Cartan sub-algebra

        p0.cx(qr[0],qr[1])
        p0.u3(a7,-halfpi,halfpi,qr[0])
        p0.h(qr[0])
        p0.u3(0,0,a9,qr[1])
        p0.cx(qr[0],qr[1])
        p0.s(qr[0])
        p0.h(qr[0])
        p0.u3(0,0,-a8,qr[1])
        p0.cx(qr[0],qr[1])
        p0.u3(-halfpi,-halfpi,halfpi,qr[0])
        p0.u3(halfpi,-halfpi,halfpi,qr[1])


# acting with K

        p0.u3(0,0,a6,qr[0])
        p0.u3(-a5,0,0,qr[0])
        p0.u3(0,0,a4,qr[0])
        p0.u3(0,0,a3,qr[1])
        p0.u3(-a2,0,0,qr[1])
        p0.u3(0,0,a1,qr[1])


        p0.measure(qr[0], cr[0])
        p0.measure(qr[1], cr[1])
        print(p0.qasm())

\end{verbatim}

When using Trotterization to evolve the states forward in time, the following Python3 code snippet, as an example, was used.  The coefficients of the terms in the Hamiltonian, as given previously, were entered as constants into the script.
\begin{verbatim}
    for ii in range(0,len(NTrotter)):
        p0=qp.get_circuit(pidtab[ii])
        ntrott = NTrotter[ii]
        print("Calculating ntrott  = ",ii," : = ",ntrott)

        for jjTT in range(0,ntrott):

            print("ii = ",ii," jjTT = ,",jjTT, "ntrott =",ntrott)

# One Trotter Step
# acting with Cartan sub-algebra to describe a1,a2,a3 = h1,h2,h3

            p0.cx(qr[0],qr[1])
            p0.u3(a1,-halfpi,halfpi,qr[0])
            p0.h(qr[0])
            p0.u3(0,0,a3,qr[1])
            p0.cx(qr[0],qr[1])
            p0.s(qr[0])
            p0.h(qr[0])
            p0.u3(0,0,-a2,qr[1])
            p0.cx(qr[0],qr[1])
            p0.u3(-halfpi,-halfpi,halfpi,qr[0])
            p0.u3(halfpi,-halfpi,halfpi,qr[1])

#  I x sigmax to describe h4

            p0.u3(a4,-halfpi,halfpi,qr[1])

#  I x sigmaz to describe h5

            p0.u3(0,0,a5,qr[1])

#   sigmaz x I to describe h6

            p0.u3(0,0,a6,qr[0])


#   sigmaz x sigmax to describe h7

            p0.h(qr[1])
            p0.cx(qr[0],qr[1])
            p0.u3(0,0,a7,qr[1])
            p0.cx(qr[0],qr[1])
            p0.h(qr[1])

# end of One Trotter Step - repeat NTrotter times

        p0.measure(qr[0], cr[0])
        p0.measure(qr[1], cr[1])
        print(p0.qasm())
\end{verbatim}

\section{Data Tables}

\begin{table}[!ht]
		\begin{tabular}{c|cccc|c}
\hline
scaled time\textbackslash\ $r$  & 1 & 3 & 5 & 7 & extrapolated\\ \hline
$\langle H \rangle$ &\text{-0.887(15)} & \text{-0.684(33)} & \text{-0.372(34)} & \text{-0.167(56)} & \text{-1.000(65)} \\
 $\langle \bar\psi \psi \rangle$ &\text{-0.3073(37)} & \text{-0.3262(39)} & \text{-0.3411(68)} & \text{-0.3515(85)} & \text{-0.296(13)} \\
 \hline
\end{tabular}
\caption{The $\langle H\rangle$ energy and $\langle \bar\psi \psi \rangle$ chiral condensate data shown in Fig.~2 of the main text.}
\end{table}

\begin{table}[!ht]
		\begin{tabular}{c|cccc|c}
\hline
scaled time\textbackslash\ $r$  & 1 & 3 & 5 & 7 & extrapolated\\ \hline
 0.1 & \text{0.0940(33)} & \text{0.1618(43)} & \text{0.2248(50)} & \text{0.2723(54)} & \text{0.055(11)} \\
 0.5 & \text{0.3299(55)} & \text{0.3163(56)} & \text{0.3185(57)} & \text{0.3093(57)} & \text{0.334(17)} \\
 0.9 & \text{0.5760(62)} & \text{0.5754(65)} & \text{0.5386(65)} & \text{0.5170(66)} & \text{0.580(20)} \\
 1.3 & \text{0.5906(65)} & \text{0.5456(65)} & \text{0.4741(64)} & \text{0.4396(64)} & \text{0.625(20)} \\
 1.7 & \text{0.4398(64)} & \text{0.3978(62)} & \text{0.3479(59)} & \text{0.3260(58)} & \text{0.470(20)} \\
 2.1 & \text{0.3333(59)} & \text{0.2850(55)} & \text{0.2828(55)} & \text{0.2921(56)} & \text{0.363(18)} \\
 2.5 & \text{0.3316(59)} & \text{0.4426(65)} & \text{0.5575(69)} & \text{0.6210(70)} & \text{0.260(19)} \\
 2.9 & \text{0.3220(58)} & \text{0.4250(64)} & \text{0.5080(68)} & \text{0.5848(69)} & \text{0.267(19)} \\
 3.3 & \text{0.2130(49)} & \text{0.2883(56)} & \text{0.3541(60)} & \text{0.3949(63)} & \text{0.167(16)} \\
 3.7 & \text{0.2434(52)} & \text{0.3253(58)} & \text{0.4085(64)} & \text{0.4820(67)} & \text{0.200(17)} \\
 4.1 & \text{0.4419(62)} & \text{0.5060(65)} & \text{0.5460(68)} & \text{0.5387(69)} & \text{0.393(20)} \\
 4.5 & \text{0.6638(62)} & \text{0.6413(65)} & \text{0.6105(67)} & \text{0.5614(67)} & \text{0.669(20)} \\
 4.9 & \text{0.6720(59)} & \text{0.6484(63)} & \text{0.6036(66)} & \text{0.5741(69)} & \text{0.687(19)} \\
 5.3 & \text{0.3374(56)} & \text{0.3499(58)} & \text{0.3624(60)} & \text{0.3886(62)} & \text{0.336(18)} \\
 5.7 & \text{0.1206(37)} & \text{0.2109(48)} & \text{0.2901(55)} & \text{0.3455(60)} & \text{0.068(13)} \\
 6.1 & \text{0.1601(43)} & \text{0.2188(49)} & \text{0.2769(55)} & \text{0.3350(59)} & \text{0.131(14)} \\
 6.5 & \text{0.3881(60)} & \text{0.3995(62)} & \text{0.4135(63)} & \text{0.4458(66)} & \text{0.389(19)} \\
 6.9 & \text{0.5735(66)} & \text{0.5690(67)} & \text{0.5538(67)} & \text{0.5360(67)} & \text{0.574(21)} \\
 7.3 & \text{0.5068(66)} & \text{0.4143(63)} & \text{0.3739(61)} & \text{0.3395(59)} & \text{0.556(20)} \\
 7.7 & \text{0.3791(61)} & \text{0.3783(61)} & \text{0.3884(62)} & \text{0.3823(62)} & \text{0.375(19)} \\
 8.1 & \text{0.3269(58)} & \text{0.3484(59)} & \text{0.3670(61)} & \text{0.3851(62)} & \text{0.316(18)} \\
 8.5 & \text{0.3061(55)} & \text{0.3346(58)} & \text{0.3651(60)} & \text{0.4014(63)} & \text{0.294(18)} \\
 8.9 & \text{0.1931(46)} & \text{0.2198(49)} & \text{0.2760(55)} & \text{0.3359(59)} & \text{0.183(15)} \\
 9.3 & \text{0.1356(40)} & \text{0.2175(49)} & \text{0.2859(55)} & \text{0.3491(60)} & \text{0.092(13)} \\
 9.7 & \text{0.2964(55)} & \text{0.3118(56)} & \text{0.3265(58)} & \text{0.3455(60)} & \text{0.290(17)} \\
 10.1 & \text{0.6204(63)} & \text{0.5665(64)} & \text{0.5351(65)} & \text{0.4920(66)} & \text{0.644(20)} \\
 10.5 & \text{0.7098(62)} & \text{0.6427(66)} & \text{0.5793(67)} & \text{0.5191(67)} & \text{0.745(20)} \\
 10.9 & \text{0.5244(62)} & \text{0.4461(63)} & \text{0.4259(64)} & \text{0.4088(63)} & \text{0.567(19)} \\
 11.3 & \text{0.2436(51)} & \text{0.2501(52)} & \text{0.2725(54)} & \text{0.3349(59)} & \text{0.254(16)} \\
 11.7 & \text{0.1594(43)} & \text{0.2159(49)} & \text{0.2860(55)} & \text{0.3434(59)} & \text{0.129(14)} \\
 12.1 & \text{0.2361(50)} & \text{0.2483(52)} & \text{0.2841(55)} & \text{0.3280(58)} & \text{0.234(16)} \\
 12.5 & \text{0.4370(62)} & \text{0.3784(60)} & \text{0.3368(58)} & \text{0.3185(58)} & \text{0.475(19)} \\
 12.9 & \text{0.4599(64)} & \text{0.3998(62)} & \text{0.3679(61)} & \text{0.3318(59)} & \text{0.489(20)} \\
 13.3 & \text{0.4105(62)} & \text{0.3800(61)} & \text{0.3816(61)} & \text{0.3800(61)} & \text{0.426(19)} \\
 13.7 & \text{0.4180(61)} & \text{0.4615(64)} & \text{0.4864(65)} & \text{0.5241(67)} & \text{0.400(19)} \\
 14.1 & \text{0.4741(64)} & \text{0.5040(66)} & \text{0.5369(68)} & \text{0.5499(69)} & \text{0.453(20)} \\
 14.5 & \text{0.3919(61)} & \text{0.4558(65)} & \text{0.4966(68)} & \text{0.5543(71)} & \text{0.364(19)} \\
 14.9 & \text{0.1828(46)} & \text{0.2734(54)} & \text{0.3651(61)} & \text{0.4619(66)} & \text{0.139(15)} \\
 15.3 & \text{0.1580(42)} & \text{0.1714(44)} & \text{0.2050(48)} & \text{0.2488(52)} & \text{0.156(14)} \\
 15.7 & \text{0.3874(59)} & \text{0.4156(61)} & \text{0.4335(63)} & \text{0.4878(67)} & \text{0.384(19)} \\
 16.1 & \text{0.6498(62)} & \text{0.5560(65)} & \text{0.4924(65)} & \text{0.4361(64)} & \text{0.700(20)} \\
 16.5 & \text{0.6677(62)} & \text{0.5594(65)} & \text{0.4781(65)} & \text{0.4220(63)} & \text{0.731(20)} \\
 16.9 & \text{0.4225(61)} & \text{0.3574(59)} & \text{0.3183(57)} & \text{0.3146(57)} & \text{0.468(19)} \\
 17.3 & \text{0.1753(45)} & \text{0.1804(45)} & \text{0.2225(50)} & \text{0.2693(54)} & \text{0.176(14)} \\
 17.7 & \text{0.2139(49)} & \text{0.3315(59)} & \text{0.4419(65)} & \text{0.5338(69)} & \text{0.149(16)} \\
 18.1 & \text{0.3479(59)} & \text{0.4248(64)} & \text{0.5003(68)} & \text{0.5613(71)} & \text{0.305(19)} \\
 18.5 & \text{0.3476(58)} & \text{0.3746(61)} & \text{0.3674(61)} & \text{0.3459(60)} & \text{0.328(18)} \\
 18.9 & \text{0.3523(60)} & \text{0.3891(62)} & \text{0.3851(62)} & \text{0.3869(62)} & \text{0.334(19)} \\
 19.3 & \text{0.4794(66)} & \text{0.5363(69)} & \text{0.5805(71)} & \text{0.5981(72)} & \text{0.442(21)} \\
 19.7 & \text{0.5473(66)} & \text{0.4968(65)} & \text{0.4630(65)} & \text{0.4474(65)} & \text{0.579(20)} \\
 \hline
\end{tabular}
\caption{The $\langle e^- e^+\rangle$ pair density data shown in Fig.~3 of the main text.}
\end{table}
\begin{table}[!ht]
		\begin{tabular}{c|cccc|c}
\hline
scaled time\textbackslash\ $r$  & 1 & 3 & 5 & 7 & extrapolated\\ \hline
 0.1 & \text{0.3003(80)} & \text{0.5185(10)} & \text{0.7313(12)} & \text{0.8938(12)} & \text{0.176(27)} \\
 0.5 & \text{0.5521(89)} & \text{0.7140(11)} & \text{0.8630(12)} & \text{1.0050(12)} & \text{0.468(29)} \\
 0.9 & \text{0.9575(92)} & \text{1.1000(99)} & \text{1.1870(10)} & \text{1.2830(11)} & \text{0.886(29)} \\
 1.3 & \text{1.1830(97)} & \text{1.2350(10)} & \text{1.2790(11)} & \text{1.3160(11)} & \text{1.150(31)} \\
 1.7 & \text{1.1700(11)} & \text{1.1680(11)} & \text{1.1610(12)} & \text{1.2380(12)} & \text{1.200(35)} \\
 2.1 & \text{1.0080(12)} & \text{0.9565(12)} & \text{1.0120(12)} & \text{1.0550(12)} & \text{1.030(38)} \\
 2.5 & \text{1.0520(14)} & \text{1.3100(14)} & \text{1.5530(14)} & \text{1.7210(14)} & \text{0.898(43)} \\
 2.9 & \text{0.9733(13)} & \text{1.2330(14)} & \text{1.4560(14)} & \text{1.6870(14)} & \text{0.844(42)} \\
 3.3 & \text{0.7448(12)} & \text{0.9188(13)} & \text{1.0550(13)} & \text{1.1300(13)} & \text{0.636(38)} \\
 3.7 & \text{0.9984(12)} & \text{1.2230(13)} & \text{1.4150(13)} & \text{1.5910(13)} & \text{0.879(38)} \\
 4.1 & \text{1.1240(11)} & \text{1.2840(11)} & \text{1.4180(11)} & \text{1.5150(11)} & \text{1.030(33)} \\
 4.5 & \text{1.1490(87)} & \text{1.2480(93)} & \text{1.3300(98)} & \text{1.3820(10)} & \text{1.090(28)} \\
 4.9 & \text{0.9450(81)} & \text{0.9739(90)} & \text{1.0620(10)} & \text{1.1370(11)} & \text{0.932(27)} \\
 5.3 & \text{0.5919(91)} & \text{0.7166(11)} & \text{0.8559(12)} & \text{1.0340(13)} & \text{0.542(30)} \\
 5.7 & \text{0.3731(89)} & \text{0.6081(11)} & \text{0.8606(12)} & \text{1.0540(13)} & \text{0.240(29)} \\
 6.1 & \text{0.4051(93)} & \text{0.6055(11)} & \text{0.7834(12)} & \text{0.9978(13)} & \text{0.314(31)} \\
 6.5 & \text{0.7404(11)} & \text{0.8303(11)} & \text{0.9920(12)} & \text{1.1540(13)} & \text{0.700(34)} \\
 6.9 & \text{1.1470(10)} & \text{1.2130(11)} & \text{1.2440(11)} & \text{1.3260(11)} & \text{1.130(33)} \\
 7.3 & \text{1.2470(11)} & \text{1.1770(12)} & \text{1.1630(12)} & \text{1.2020(12)} & \text{1.300(35)} \\
 7.7 & \text{1.2310(12)} & \text{1.1850(12)} & \text{1.2120(12)} & \text{1.2350(12)} & \text{1.260(37)} \\
 8.1 & \text{1.0190(12)} & \text{1.0390(12)} & \text{1.0970(12)} & \text{1.1640(12)} & \text{1.010(37)} \\
 8.5 & \text{0.7679(11)} & \text{0.8674(11)} & \text{0.9961(12)} & \text{1.1240(13)} & \text{0.720(34)} \\
 8.9 & \text{0.5951(10)} & \text{0.7538(11)} & \text{0.9345(12)} & \text{1.1110(13)} & \text{0.516(33)} \\
 9.3 & \text{0.5836(11)} & \text{0.7115(11)} & \text{0.8779(12)} & \text{1.0380(13)} & \text{0.520(34)} \\
 9.7 & \text{0.7531(11)} & \text{0.8147(11)} & \text{0.9195(12)} & \text{1.0820(12)} & \text{0.743(34)} \\
 10.1 & \text{0.9981(93)} & \text{0.9983(10)} & \text{1.0550(11)} & \text{1.1450(11)} & \text{1.010(30)} \\
 10.5 & \text{1.1090(86)} & \text{1.1540(98)} & \text{1.1920(11)} & \text{1.2590(11)} & \text{1.090(28)} \\
 10.9 & \text{0.9039(93)} & \text{0.9494(11)} & \text{1.0600(12)} & \text{1.1300(12)} & \text{0.874(31)} \\
 11.3 & \text{0.6156(10)} & \text{0.7219(12)} & \text{0.8358(12)} & \text{1.0040(13)} & \text{0.580(34)} \\
 11.7 & \text{0.5386(11)} & \text{0.7541(13)} & \text{1.0310(13)} & \text{1.2210(14)} & \text{0.408(36)} \\
 12.1 & \text{0.5654(10)} & \text{0.6820(11)} & \text{0.8614(12)} & \text{1.0110(12)} & \text{0.502(33)} \\
 12.5 & \text{0.8863(11)} & \text{0.9171(11)} & \text{0.9903(12)} & \text{1.0630(12)} & \text{0.873(34)} \\
 12.9 & \text{1.2130(11)} & \text{1.2680(11)} & \text{1.3280(12)} & \text{1.3400(12)} & \text{1.170(34)} \\
 13.3 & \text{1.2630(11)} & \text{1.1990(11)} & \text{1.1680(12)} & \text{1.1620(12)} & \text{1.300(34)} \\
 13.7 & \text{1.2110(10)} & \text{1.2440(11)} & \text{1.2270(11)} & \text{1.2440(11)} & \text{1.200(33)} \\
 14.1 & \text{1.0440(11)} & \text{1.1380(11)} & \text{1.2150(12)} & \text{1.2620(12)} & \text{0.986(34)} \\
 14.5 & \text{0.8321(11)} & \text{1.0300(12)} & \text{1.1960(13)} & \text{1.3800(13)} & \text{0.737(36)} \\
 14.9 & \text{0.5653(11)} & \text{0.8521(13)} & \text{1.0670(13)} & \text{1.3090(14)} & \text{0.425(36)} \\
 15.3 & \text{0.4758(98)} & \text{0.5831(11)} & \text{0.7408(12)} & \text{0.8985(12)} & \text{0.426(32)} \\
 15.7 & \text{0.7096(98)} & \text{0.7891(11)} & \text{0.8935(11)} & \text{1.0340(12)} & \text{0.683(31)} \\
 16.1 & \text{1.0650(89)} & \text{1.1390(10)} & \text{1.1800(11)} & \text{1.3010(12)} & \text{1.050(29)} \\
 16.5 & \text{1.0970(88)} & \text{1.1380(10)} & \text{1.1960(11)} & \text{1.2610(11)} & \text{1.080(29)} \\
 16.9 & \text{0.8708(10)} & \text{0.9229(11)} & \text{1.0070(12)} & \text{1.1070(12)} & \text{0.852(33)} \\
 17.3 & \text{0.5965(11)} & \text{0.6764(11)} & \text{0.8040(12)} & \text{0.9610(12)} & \text{0.569(34)} \\
 17.7 & \text{0.6991(12)} & \text{1.0170(13)} & \text{1.2600(14)} & \text{1.4610(14)} & \text{0.522(38)} \\
 18.1 & \text{0.8359(11)} & \text{1.0690(12)} & \text{1.2380(12)} & \text{1.3800(12)} & \text{0.707(36)} \\
 18.5 & \text{0.9859(11)} & \text{1.1360(12)} & \text{1.2500(12)} & \text{1.3120(12)} & \text{0.892(35)} \\
 18.9 & \text{1.2700(12)} & \text{1.4550(12)} & \text{1.5280(12)} & \text{1.5660(12)} & \text{1.160(36)} \\
 19.3 & \text{1.4180(11)} & \text{1.5030(11)} & \text{1.5370(11)} & \text{1.5640(12)} & \text{1.370(35)} \\
 19.7 & \text{1.1560(10)} & \text{1.1790(11)} & \text{1.2250(11)} & \text{1.2560(12)} & \text{1.140(33)} \\
 \hline
\end{tabular}
\caption{The $\langle E^2\rangle$ energy in the electric field data shown in Fig.~3 of the main text.}
\end{table}

\begin{table}[!ht]
		\begin{tabular}{c|cccc|c||c|c}
\hline
scaled time \textbackslash\ $r$  & 1 & 3 & 5 & 7 & extrapolated & scaled time \textbackslash\ $r$ & 1\\ \hline
0.1 & \text{0.0710(36)} & \text{0.1178(46)} & \text{0.1780(54)} & \text{0.2632(62)} & \text{0.056(12)} & 0.25 & \text{0.1394(40)} \\
 0.2 & \text{0.1390(49)} & \text{0.2040(57)} & \text{0.3302(67)} & \text{0.4426(70)} & \text{0.108(16)} & 0.5 & \text{0.3349(56)} \\
 0.3 & \text{0.2208(59)} & \text{0.3086(65)} & \text{0.4432(70)} & \text{0.5060(71)} & \text{0.158(19)} & 0.75 & \text{0.5503(64)} \\
 0.4 & \text{0.3038(65)} & \text{0.3818(69)} & \text{0.4888(71)} & \text{0.4932(71)} & \text{0.233(21)} & 1. & \text{0.6490(66)} \\
 0.5 & \text{0.3734(68)} & \text{0.4422(70)} & \text{0.4992(71)} & \text{0.4834(71)} & \text{0.314(21)} & 1.25 & \text{0.6688(69)} \\
 0.6 & \text{0.4446(70)} & \text{0.4932(71)} & \text{0.4992(71)} & \text{0.4584(70)} & \text{0.403(22)} & 1.5 & \text{0.6351(71)} \\
 0.7 & \text{0.4920(71)} & \text{0.4798(71)} & \text{0.4886(71)} & \text{0.4592(70)} & - & 1.75 & \text{0.5774(71)} \\
 0.8 & \text{0.5298(71)} & - & - & - & - & 2. & \text{0.5404(70)} \\
 0.9 & \text{0.5436(70)} & - & - & - & - & - & - \\
 1. & \text{0.5328(71)} & - & - & - & - & - & - \\
 1.1 & \text{0.5488(70)} & - & - & - & - & - & - \\
 1.2 & \text{0.5400(70)} & - & - & - & - & - & - \\
 1.3 & \text{0.5180(71)} & - & - & - & - & - & - \\
 1.4 & \text{0.5166(71)} & - & - & - & - & - & - \\
 1.5 & \text{0.5104(71)} & - & - & - & - & - & - \\
 1.6 & \text{0.5174(71)} & - & - & - & - & - & - \\
 1.7 & \text{0.5150(71)} & - & - & - & - & - & - \\
 1.8 & \text{0.5276(71)} & - & - & - & - & - & - \\
 1.9 & \text{0.5132(71)} & - & - & - & - & - & - \\
 2. & \text{0.5154(71)} & - & - & - & - & - & - \\
 \hline
\end{tabular}
\caption{The $\langle e^-e^+\rangle$ pair density data shown in Fig.~4 of the main text.}
\end{table}

\begin{table}[!ht]
\scalebox{0.8}{
		\begin{tabular}{c|cccc}
\hline
scaled time  & 3-CNOT (sim) & 3-CNOT (ibmqx2) & 6-CNOT (sim) & 6-CNOT (ibmqx2)\\ \hline
0.1 & \text{0.9873(35)} & \text{0.9023(66)} & \text{0.9945(22)} & \text{0.8320(24)} \\
 0.5 & \text{0.8477(11)} & \text{0.8018(88)} & \text{0.8440(10)} & \text{0.7705(24)} \\
 0.9 & \text{0.6494(15)} & \text{0.6411(11)} & \text{0.6370(16)} & \text{0.6084(25)} \\
 1.3 & \text{0.5049(16)} & \text{0.5029(11)} & \text{0.5170(19)} & \text{0.4668(26)} \\
 1.7 & \text{0.5498(16)} & \text{0.4844(11)} & \text{0.5015(19)} & \text{0.4814(27)} \\
 2.1 & \text{0.5420(16)} & \text{0.5352(11)} & \text{0.5355(20)} & \text{0.5703(28)} \\
 2.5 & \text{0.6318(15)} & \text{0.6191(11)} & \text{0.5975(20)} & \text{0.6045(29)} \\
 2.9 & \text{0.6436(15)} & \text{0.6128(11)} & \text{0.6185(20)} & \text{0.6045(29)} \\
 3.3 & \text{0.6465(15)} & \text{0.6460(11)} & \text{0.6405(19)} & \text{0.5264(27)} \\
 3.7 & \text{0.7178(14)} & \text{0.6860(10)} & \text{0.7080(17)} & \text{0.6348(27)} \\
 4.1 & \text{0.8652(11)} & \text{0.8496(79)} & \text{0.8835(12)} & \text{0.7490(24)} \\
 4.5 & \text{0.9883(34)} & \text{0.8979(67)} & \text{0.9900(85)} & \text{0.8164(23)} \\
 4.9 & \text{0.9170(86)} & \text{0.8589(77)} & \text{0.9320(11)} & \text{0.7617(25)} \\
 5.3 & \text{0.7041(14)} & \text{0.6812(10)} & \text{0.6900(15)} & \text{0.6357(26)} \\
 5.7 & \text{0.4482(16)} & \text{0.5215(11)} & \text{0.4525(17)} & \text{0.4307(25)} \\
 6.1 & \text{0.3965(15)} & \text{0.4058(11)} & \text{0.3870(18)} & \text{0.4102(25)} \\
 6.5 & \text{0.4463(16)} & \text{0.4258(11)} & \text{0.4830(19)} & \text{0.5088(27)} \\
 6.9 & \text{0.5957(15)} & \text{0.6406(11)} & \text{0.5815(20)} & \text{0.5693(28)} \\
 7.3 & \text{0.7041(14)} & \text{0.6650(10)} & \text{0.7040(21)} & \text{0.5732(28)} \\
 7.7 & \text{0.7256(14)} & \text{0.7510(96)} & \text{0.7410(19)} & \text{0.5947(27)} \\
 8.1 & \text{0.7686(13)} & \text{0.7393(97)} & \text{0.7780(17)} & \text{0.6660(27)} \\
 8.5 & \text{0.8213(12)} & \text{0.7690(93)} & \text{0.8330(16)} & \text{0.7129(27)} \\
 8.9 & \text{0.9268(81)} & \text{0.8477(79)} & \text{0.9210(13)} & \text{0.7607(27)} \\
 9.3 & \text{0.9521(67)} & \text{0.8818(71)} & \text{0.9635(14)} & \text{0.8105(27)} \\
 9.7 & \text{0.8047(12)} & \text{0.7690(93)} & \text{0.8090(17)} & \text{0.6719(27)} \\
 10.1 & \text{0.5039(16)} & \text{0.5283(11)} & \text{0.5325(17)} & \text{0.5547(27)} \\
 10.5 & \text{0.3232(15)} & \text{0.3677(11)} & \text{0.3255(16)} & \text{0.3730(24)} \\
 10.9 & \text{0.3438(15)} & \text{0.3706(11)} & \text{0.3420(17)} & \text{0.3682(24)} \\
 11.3 & \text{0.5410(16)} & \text{0.4634(11)} & \text{0.4980(19)} & \text{0.4688(26)} \\
 11.7 & \text{0.6963(14)} & \text{0.6016(11)} & \text{0.6915(20)} & \text{0.6377(28)} \\
 12.1 & \text{0.8135(12)} & \text{0.7930(90)} & \text{0.8020(18)} & \text{0.7139(28)} \\
 12.5 & \text{0.8477(11)} & \text{0.8110(87)} & \text{0.8455(16)} & \text{0.7080(28)} \\
 12.9 & \text{0.8193(12)} & \text{0.7739(92)} & \text{0.8030(15)} & \text{0.6768(26)} \\
 13.3 & \text{0.8320(12)} & \text{0.7754(92)} & \text{0.8210(15)} & \text{0.6943(26)} \\
 13.7 & \text{0.8760(10)} & \text{0.7998(88)} & \text{0.8880(18)} & \text{0.7373(28)} \\
 14.1 & \text{0.8584(11)} & \text{0.8081(87)} & \text{0.8640(20)} & \text{0.7207(28)} \\
 14.5 & \text{0.6836(15)} & \text{0.6978(10)} & \text{0.6660(20)} & \text{0.6709(29)} \\
 14.9 & \text{0.4219(15)} & \text{0.4199(11)} & \text{0.4280(18)} & \text{0.4033(25)} \\
 15.3 & \text{0.2754(14)} & \text{0.4204(11)} & \text{0.2855(15)} & \text{0.3486(23)} \\
 15.7 & \text{0.3818(15)} & \text{0.4404(11)} & \text{0.3820(17)} & \text{0.4570(26)} \\
 16.1 & \text{0.6250(15)} & \text{0.6011(11)} & \text{0.6175(18)} & \text{0.5869(27)} \\
 16.5 & \text{0.8252(12)} & \text{0.8003(88)} & \text{0.8220(17)} & \text{0.7266(26)} \\
 16.9 & \text{0.9150(87)} & \text{0.8384(81)} & \text{0.9220(14)} & \text{0.7061(26)} \\
 17.3 & \text{0.8906(98)} & \text{0.8125(86)} & \text{0.8860(12)} & \text{0.7061(26)} \\
 17.7 & \text{0.7920(13)} & \text{0.7612(94)} & \text{0.7890(15)} & \text{0.6553(26)} \\
 18.1 & \text{0.7471(14)} & \text{0.6924(10)} & \text{0.7380(19)} & \text{0.6318(27)} \\
 18.5 & \text{0.7939(13)} & \text{0.7256(99)} & \text{0.7865(20)} & \text{0.6836(29)} \\
 18.9 & \text{0.7285(14)} & \text{0.7490(96)} & \text{0.7500(21)} & \text{0.7031(30)} \\
 19.3 & \text{0.6289(15)} & \text{0.5884(11)} & \text{0.6170(21)} & \text{0.5908(29)} \\
 19.7 & \text{0.3994(15)} & \text{0.4468(11)} & \text{0.3900(18)} & \text{0.4229(26)} \\
 20.1 & \text{0.3564(15)} & \text{0.3994(11)} & \text{0.3555(17)} & \text{0.4131(25)} \\
 20.5 & \text{0.5117(16)} & \text{0.5137(11)} & \text{0.5105(17)} & \text{0.5205(26)} \\
 20.9 & \text{0.7949(13)} & \text{0.7539(95)} & \text{0.7625(15)} & \text{0.6709(27)} \\
 21.3 & \text{0.9502(68)} & \text{0.8555(78)} & \text{0.9375(11)} & \text{0.7744(25)} \\
 21.7 & \text{0.9473(70)} & \text{0.8594(77)} & \text{0.9595(76)} & \text{0.7559(23)} \\
 22.1 & \text{0.8154(12)} & \text{0.7769(92)} & \text{0.8330(12)} & \text{0.6904(25)} \\
 22.5 & \text{0.6523(15)} & \text{0.6885(10)} & \text{0.6665(15)} & \text{0.5840(24)} \\
 22.9 & \text{0.6016(15)} & \text{0.6172(11)} & \text{0.6115(19)} & \text{0.5420(26)} \\
 23.3 & \text{0.6523(15)} & \text{0.6318(11)} & \text{0.6615(21)} & \text{0.6211(29)} \\
 23.7 & \text{0.6904(14)} & \text{0.6460(11)} & \text{0.6685(21)} & \text{0.6143(29)} \\
 24.1 & \text{0.5684(15)} & \text{0.5820(11)} & \text{0.5740(20)} & \text{0.5391(28)} \\
 24.5 & \text{0.4473(16)} & \text{0.4272(11)} & \text{0.4550(19)} & \text{0.4561(26)} \\
 24.9 & \text{0.4668(16)} & \text{0.5063(11)} & \text{0.4855(17)} & \text{0.5039(25)} \\
 \hline
\end{tabular}}
\caption{The $\langle e^-e^+\rangle$ pair density data shown in Fig.~7.}
\end{table}

\begin{table}[!ht]
		\begin{tabular}{c|cccc|c}
\hline
scaled time\textbackslash\ $r$  & 1 & 3 & 5 & 7 & extrapolated\\ \hline
 0.1 & \text{0.0898(32)} & \text{0.1983(45)} & \text{0.3176(52)} & \text{0.4386(55)} & \text{0.037(11)} \\
 0.5 & \text{0.2979(51)} & \text{0.2803(5)} & \text{0.2613(49)} & \text{0.2574(49)} & \text{0.312(16)} \\
 0.9 & \text{0.5331(56)} & \text{0.4826(56)} & \text{0.4405(56)} & \text{0.4110(55)} & \text{0.563(17)} \\
 1.3 & \text{0.5299(56)} & \text{0.4439(56)} & \text{0.3820(54)} & \text{0.3408(53)} & \text{0.581(17)} \\
 1.7 & \text{0.3840(54)} & \text{0.3364(53)} & \text{0.2946(51)} & \text{0.2804(5)} & \text{0.418(17)} \\
 2.1 & \text{0.3646(54)} & \text{0.4558(56)} & \text{0.5358(56)} & \text{0.5674(55)} & \text{0.303(17)} \\
 2.5 & \text{0.3218(52)} & \text{0.4089(55)} & \text{0.4725(56)} & \text{0.4904(56)} & \text{0.262(17)} \\
 2.9 & \text{0.2745(5)} & \text{0.2976(51)} & \text{0.3110(52)} & \text{0.2910(51)} & \text{0.251(16)} \\
 3.3 & \text{0.1541(4)} & \text{0.1836(43)} & \text{0.2025(45)} & \text{0.2241(47)} & \text{0.140(13)} \\
 3.7 & \text{0.2001(45)} & \text{0.2660(49)} & \text{0.2810(5)} & \text{0.2860(51)} & \text{0.161(14)} \\
 4.1 & \text{0.4467(56)} & \text{0.4704(56)} & \text{0.4793(56)} & \text{0.5001(56)} & \text{0.438(17)} \\
 4.5 & \text{0.6786(52)} & \text{0.6550(53)} & \text{0.6273(54)} & \text{0.6081(55)} & \text{0.693(16)} \\
 4.9 & \text{0.6384(54)} & \text{0.5699(55)} & \text{0.5285(56)} & \text{0.5061(56)} & \text{0.680(17)} \\
 5.3 & \text{0.3280(52)} & \text{0.3863(54)} & \text{0.4346(55)} & \text{0.4691(56)} & \text{0.294(17)} \\
 5.7 & \text{0.0863(31)} & \text{0.1978(45)} & \text{0.3033(51)} & \text{0.3910(55)} & \text{0.025(11)} \\
 6.1 & \text{0.1740(42)} & \text{0.2253(47)} & \text{0.2305(47)} & \text{0.2199(46)} & \text{0.141(13)} \\
 6.5 & \text{0.4066(55)} & \text{0.4713(56)} & \text{0.4973(56)} & \text{0.5256(56)} & \text{0.374(17)} \\
 6.9 & \text{0.5708(55)} & \text{0.6149(54)} & \text{0.6304(54)} & \text{0.6363(54)} & \text{0.544(17)} \\
 7.3 & \text{0.4770(56)} & \text{0.4509(56)} & \text{0.4265(55)} & \text{0.4031(55)} & \text{0.490(17)} \\
 7.7 & \text{0.3564(54)} & \text{0.3869(54)} & \text{0.4100(55)} & \text{0.4119(55)} & \text{0.334(17)} \\
 8.1 & \text{0.2973(51)} & \text{0.3559(54)} & \text{0.3893(55)} & \text{0.4061(55)} & \text{0.261(16)} \\
 8.5 & \text{0.2869(51)} & \text{0.2536(49)} & \text{0.2501(48)} & \text{0.2918(51)} & \text{0.320(16)} \\
 8.9 & \text{0.1894(44)} & \text{0.2043(45)} & \text{0.2399(48)} & \text{0.2901(51)} & \text{0.188(14)} \\
 9.3 & \text{0.1114(35)} & \text{0.1875(44)} & \text{0.3055(52)} & \text{0.4228(55)} & \text{0.077(12)} \\
 9.7 & \text{0.3490(53)} & \text{0.4104(55)} & \text{0.4721(56)} & \text{0.5041(56)} & \text{0.309(17)} \\
 10.1 & \text{0.5988(55)} & \text{0.5140(56)} & \text{0.4241(55)} & \text{0.3495(53)} & \text{0.646(17)} \\
\hline
		\end{tabular}
\caption{
The $\langle e^-e^+ \rangle$ pair density
data shown in Fig.~10.
}
\label{tab:CNOTextrapPairs}
\end{table}

\begin{table}[!ht]
		\begin{tabular}{c|cccc|c}
\hline
operator\textbackslash\ $r$  & 1 & 3 & 5 & 7 & extrapolated\\ \hline
 $\sigma _y\text{ $\otimes $ }\sigma _y $& \text{-0.105(15)} & \text{-0.097(11)} & \text{-0.074(16)} & \text{-0.036(10)} &
   \text{-0.104(46)} \\
 $ \text{I $\otimes $ }\sigma _x$ & \text{-0.893(09)} & \text{-0.733(19)} & \text{-0.518(15)} & \text{-0.351(37)} &
   \text{-0.980(39)} \\
 $\sigma _z\text{ $\otimes $ }\sigma _z $& \text{0.311(11)} & \text{0.489(12)} & \text{0.631(18)} & \text{0.687(22)} &
   \text{0.196(37)} \\
 $\sigma _z\text{ $\otimes $ I} $& \text{0.918(06)} & \text{0.816(10)} & \text{0.734(13)} & \text{0.719(21)} & \text{0.988(25)} \\
 $\sigma _z\text{ $\otimes $ }\sigma _x$ & \text{-0.867(11)} & \text{-0.668(20)} & \text{-0.467(17)} & \text{-0.293(29)} &
   \text{-0.974(40)} \\
$\text{I $\otimes $ }\sigma _z $& \text{0.351(11)} & \text{0.540(13)} & \text{0.672(15)} & \text{0.731(17)} & \text{0.230(36)} \\
 $\sigma _x\text{ $\otimes $ }\sigma _x $& \text{-0.137(13)} & \text{-0.135(12)} & \text{-0.051(24)} & \text{-0.035(22)} &
   \text{-0.138(45)} \\
\hline
		\end{tabular}
\caption{
The data associated with $\langle \mathcal{O} \rangle$
as a function of
 the noise parameter in Fig.~11.
}
\label{tab:rCNOTvar}
\end{table}

\begin{table}[!ht]
		\begin{tabular}{c|cc}
\hline
$r$\ \textbackslash\   scaled time  & 2.4 & 6.4 \\ \hline
 1 & \text{0.3899(55)} & \text{0.7646(47)} \\
 3 & \text{0.2950(51)} & \text{0.6994(51)} \\
 5 & \text{0.2281(47)} & \text{0.6671(53)} \\
 7 & \text{0.2076(45)} & \text{0.6338(54)} \\
 9 & \text{0.2165(92)} & \text{0.565(11)} \\
 11 & \text{0.2435(68)} & \text{0.4998(79)} \\
 13 & \text{0.2775(71)} & \text{0.4478(79)} \\
 15 & \text{0.2963(72)} & \text{0.3883(77)} \\
 17 & \text{0.2813(71)} & \text{0.3710(76)} \\
 19 & \text{0.2323(67)} & \text{0.2888(72)} \\
 21 & \text{0.2493(68)} & \text{0.2838(71)} \\
 23 & \text{0.2038(64)} & \text{0.2370(67)} \\
 25 & \text{0.2103(64)} & \text{0.2543(69)} \\
\hline
		\end{tabular}
\caption{
The data associated with $\langle 0 \rangle$
as a function of
the number of the noise parameter determined at two times
that is shown in Fig.~12.
}
\label{tab:rCNOTextrap0Pairs}
\end{table}

\FloatBarrier
\end{appendix}

%%%%%%%%%%%%%%%%%%%%%%%%%%%%%%%%%%%%%%%%%%%%%%%%%%
\bibliography{QFT}
\end{document}